\documentclass[article,11pt]{article}
\usepackage[includeheadfoot,
            marginratio={1:1,2:3}, 
            width=460pt,
            height=688pt,]{geometry}
            \usepackage{tikz}
\usepackage[utf8]{inputenc}
\usepackage{amsmath}
\usepackage{amsfonts}
\usepackage{amssymb}
\usepackage{amsthm}
\usepackage{graphicx}
\usepackage{booktabs}
\usepackage{verbatim}
\usepackage{empheq}
\usepackage{paralist}
\usepackage{cite}
\usepackage[hidelinks]{hyperref}
\usepackage{xcolor}
\usepackage{tensor}
\usepackage{adjustbox}
\usepackage{subcaption}
\usepackage{braket}
\usepackage{tikz}
\usepackage{tikz-cd}
\usepackage{yfonts}
\usepackage{enumitem}
\usepackage{physics}
\usepackage{slashed}
\usepackage{mdframed}
\usepackage{simpler-wick}
\usepackage{hhline}
\usepackage{listings}
\usepackage{mathtools}
\usepackage{xcolor}
\usetikzlibrary{decorations.pathreplacing,calc}

\newcommand{\nc}{\newcommand}
\nc{\lb}{\llbracket}
\nc{\rb}{\rrbracket}
\nc{\gl}{\llbracket}
\nc{\gr}{\rrbracket}
\nc{\bbR}{\mathbb{R}}
\nc{\bbC}{\mathbb{C}}
\nc{\bbZ}{\mathbb{Z}}
\nc{\cO}{\mathcal{O}}
\nc{\cS}{\mathcal{S}}
\nc{\cM}{\mathcal{M}}
\nc{\cT}{\mathcal{T}}
\nc{\cX}{\mathcal{X}}
\nc{\cQ}{\mathcal{Q}}
\nc{\cA}{\mathcal{A}}
\nc{\cD}{\mathcal{D}}
\nc{\cL}{\mathcal{L}}
\nc{\cC}{\mathcal{C}}
\nc{\cG}{\mathcal{G}}
\nc{\cF}{\mathcal{F}}
\nc{\cI}{\mathcal{I}}
\nc{\cN}{\mathcal{N}}
\nc{\pd}{\partial}
\nc{\la}{\lambda}

\newcommand\beq{\begin{equation}}
\newcommand\eeq{\end{equation}}
\nc{\del}{\partial}
\nc{\tri}{\hspace{-3.5pt}\vartriangle\hspace{-3.5pt}}
\nc{\blacktri}{\blacktriangle}
\nc{\eq}[1]{\begin{equation}
                     \begin{split} #1 \end{split}
                     \end{equation}}
\nc{\ul}{\underline}
\nc{\ov}{\overline}
\nc{\fa}{\hat}
\nc{\fb}{\MakeUppercase}
\nc{\fc}{\tilde }
\nc{\Lie}{{\cal L}} 
\nc{\lambdabar}{{\mkern0.75mu\mathchar '26\mkern -9.75mu\lambda}}

\newmuskip\pFqmuskip

\newcommand*\pFq[7][8]{
  \begingroup 
  \pFqmuskip=#1mu\relax
  \mathchardef\normalcomma=\mathcode`,
  \mathcode`\,=\string"8000
  \begingroup\lccode`\~=`\,
  \lowercase{\endgroup\let~}\pFqcomma
  {}_{#2}{#3}_{#4}{\left[\left.\genfrac..{0pt}{}{#5}{#6}\right|#7\right]}
  \endgroup
}
\newcommand{\pFqcomma}{{\normalcomma}\mskip\pFqmuskip}

\allowdisplaybreaks[2]

\theoremstyle{definition}

\begin{document}

\vspace*{1.5cm}
\begin{center}
{\LARGE
\textbf{Fisher information in the flux landscape}}\\[.4cm]
{\Large
}

\end{center}

\vspace{0.35cm}
\begin{center}
 {\hspace{-0.05cm}\bf Martin Carrascal$^{1}$, Radu Tatar$^1$}
\end{center}

\vspace{.25cm}
\begin{center} 
\emph{
$^1$\, Department of Mathematical Sciences, University of Liverpool,\\ 
Liverpool, L69 7ZL, United Kingdom}\\
\end{center}

\vspace{2.5cm}


\begin{abstract}
\noindent
We quantify the fine-tuning of the flux superpotential in Type IIB Calabi–Yau orientifold compactifications using the Fisher Information Metric. Treating the background fluxes as model parameters and the magnitude of the superpotential as the physical observable, we show that the Fisher matrix admits a unique non-zero eigenvalue that bounds the superpotential from above and factorises into axio-dilaton, complex structure components and a bounded factor depending on both moduli and the phase of the superpotential. In the canonical duality frame and over the full flux lattice, this eigenvalue grows linearly with the inverse string coupling and cubically with the complex structure saxions. Specialising to a two modulus orientifold of the degree 18 hypersurface in $\mathbb{P}^4(1,1,1,6,9)$ at large complex structure, we combine the flux integrality with the D3-brane tadpole condition to show that several flux components, including the one driving this growth, are forced to vanish across deep moduli space patches, truncating the twelve-dimensional flux lattice to an eight-dimensional sublattice. The Fisher eigenvalue restricted to these admissible flux directions either approaches a constant or grows at most linearly, much more slowly than over the full lattice.
\bigskip
\end{abstract}

\clearpage

\tableofcontents


\newpage

\parskip=.2cm

\section{Introduction}
One of the most intriguing questions in modern physics is how to understand the nature of certain parameters that appear ubiquitously in our theories of the universe. We say that a parameter is natural if, when set to zero, the system acquires a new symmetry \cite{tHooft:1979rat}. As an example, the small values of fermion masses in the Standard Model (SM) compared to high-energy scales such as the Planck mass \(M_{\text{pl}}\) can be naturally explained by the emergence of a chiral symmetry when those masses are set to zero. By contrast, there are still many SM parameters that are considered unnatural, as there is no known symmetry underlying their values. The most prominent example in particle physics is the Higgs mass, whose bare value squared receives radiative corrections of the order of \(M_{\text{pl}}^{2}\sim10^{38}\text{ GeV}^{2}\), meaning that an incredible degree of fine-tuned cancellation between the bare mass and these corrections is required to account for the measured value of \(m_{H}\sim 125 \text{ GeV}\). This naturalness problem is known as the hierarchy problem \cite{Witten:1981nf, Dimopoulos:1981zb, Susskind:1978ms, Arkani-Hamed:1998jmv}. A second long-standing puzzle concerns the cosmological constant: in order to reproduce the observed value $\rho_{\Lambda} \sim 10^{-120} M_{\text{pl}}^{4}$, the bare vacuum energy, alongside classical symmetry-breaking contributions such as the electroweak (EW) Higgs potential minimum, must cancel against enormous quantum loop corrections \cite{Weinberg:1988cp,Martin:2012bt,Burgess:2013ara}.

A relatively recent approach to addressing these naturalness puzzles comes from the Swampland Program \cite{Vafa:2005ui}, see \cite{Palti:2019pca, vanBeest:2021lhn} for reviews. In short, the consistency conditions imposed by quantum gravity might explain the smallness of parameters that naively appear fine-tuned in the corresponding EFTs \cite{Ibanez:2017oqr, Ibanez:2017kvh, Montero:2022prj}. Despite this progress, there is still no consensus on how to quantify naturalness, i.e. on a measure capturing the degree of fine-tuning of the parameters from which physical observables are derived \cite{Giudice:2008bi}. 

A novel proposal was put forward in~\cite{Halverson:2026cxe}, where the authors motivate a naturalness measure based on information theory. This quantity encodes the sensitivity of physical observables to changes in the model's parameters. To this end, a probability distribution over observable space is associated with every point in parameter space, and the distance between different distributions is subsequently computed. Under certain assumptions, the metric associated with this distance turns out to be the Fisher Information Metric (FIM), whose eigenvalues determine how natural or unnatural variations in parameter space are. The authors investigate a variety of examples in which this idea is put into practice. 

In this work we consider a further application of the proposal in \cite{Halverson:2026cxe}. In particular, we study the sensitivity of the flux superpotential to variations in flux space in the context of Type IIB compactifications on a Calabi-Yau orientifold. The main motivation for studying the superpotential as an observable is the requirement that it should be sufficiently small to give rise to realistic flux vacua in moduli stabilisation scenarios such as KKLT \cite{Kachru:2003aw, Demirtas:2019sip, Broeckel:2020fdz}. Achieving such small values requires a somewhat miraculous cancellation among flux-dependent quantities, which we examine by studying the behaviour of the unique non-zero eigenvalue of the fine-tuning matrix.

The paper is organised as follows. In section \ref{sec:setup}, we begin by reviewing general aspects of the geometry associated with Type IIB compactifications on a Calabi-Yau threefold. This section concludes with a description of a specific geometry that will be exploited in sections \ref{sec:fine-tuning-and-distance} and \ref{sec:fine-tuning-and-string-coupling}. In section \ref{sec:fisher}, we begin by briefly reviewing the framework of \cite{Halverson:2026cxe} and apply this idea to the flux landscape. In subsection \ref{sec:fine-tuning-and-distance}, we study how the integrality of the fluxes, subject to the tadpole bound, can be used to reduce the effective dimension of the flux lattice across certain regions of moduli space. Section 3 concludes with the asymptotic behaviour of the Fisher eigenvalue at weak coupling and large complex structure. We show that the growth found over the full flux lattice is removed once the tadpole bound is imposed, illustrating this on the special flux branch of the geometry in section \ref{sec:specialisation-geoemtry}. Finally, in section \ref{sec:conclusions}, we present our conclusions and outlook.

\section{Moduli space geometry, periods and fluxes}
\label{sec:setup}

In this section, we review the geometric ingredients underlying the complex structure moduli space of Type IIB compactifications on Calabi-Yau manifolds and introduce a specific compactification with $h^{2,1}=2$ that will be exploited from section \ref{sec:fine-tuning-and-distance} onwards.

\subsection{General features}\label{subsec:periods}

Let us consider Type IIB string theory compactified on a Calabi--Yau threefold $Y$.
The complex structure deformations of $Y$ span a special K\"ahler manifold
$\mathcal{M}_{\mathrm{cs}}$ of complex dimension $h^{2,1}$.
Choosing a real symplectic basis
$(\alpha_I,\beta^I)$ of $H^3(Y,\mathbb{Z})$ satisfying
\begin{equation}
\int_{Y}\alpha_I\wedge \beta^J=\delta_I^{\,J},\qquad I=0,\dots,h^{2,1},
\end{equation}
one defines the periods of $Y$ as the holomorphic quantities that arise from integrating the holomorphic 3-form $\Omega(z)$ over the basis of 3-cycles dual to the aforementioned symplectic one. These can be written as
\begin{equation}
X^I=\int_{Y}\Omega\wedge\beta^I,
\qquad
\mathcal{F}_I=\int_{Y}\Omega\wedge\alpha_I, \qquad \mathbf{\Pi} = \begin{pmatrix} 
 
\mathcal{F}_{I}\\ 
X^{I} \\ 
\end{pmatrix},
\label{eq:general_periods}
\end{equation}
with $\mathbf{\Pi}$ being the symplectic period vector. Since $\Omega$ is defined only up to an overall holomorphic rescaling, the physical
complex structure coordinates are given by the special coordinates
\begin{equation}
z^i=\frac{X^i}{X^0},
\qquad
i=1,\dots,h^{2,1}.
\end{equation}
By means of mirror symmetry, the complex structure moduli space of $Y$ is identified with the
K\"ahler moduli space of the mirror threefold $\tilde Y$. In particular, near large
complex structure (LCS) the prepotential takes the standard form \cite{Candelas:1990pi, Hosono:1993qy,Grimm:2024fip}
\begin{equation}\label{eq:general_prepotential}
\mathcal{F}(z)=
-\frac{1}{6}\kappa_{ijk}z^iz^jz^k
+\frac{1}{2}a_{ij}z^iz^j
+b_i z^i
-\frac{\chi\,\zeta(3)}{2(2\pi i)^3}
+\mathcal{O}(e^{2\pi i z}),
\end{equation}
where $\kappa_{ijk}$ are the triple intersection numbers, which together with
$a_{ij},b_i$ and $\chi$ represent topological data of the mirror $\tilde Y$. More explicitly, they are obtained by integration of Chern classes and basis generators $\tilde{\omega}_{i}$ of $H^{1,1}(\tilde{Y},\mathbb{Z})$ as follows
\begin{equation}\label{eq:mirror-data}
    \kappa_{ijk}=\int_{\tilde{Y}}\tilde{\omega}_{i}\wedge \tilde{\omega}_{j}\wedge \tilde{\omega}_{k},\quad b_{i}=\frac{1}{4!}\int_{\tilde{Y}}c_{2}(\tilde{Y})\wedge \tilde{\omega}_{i},\quad \chi=\int_{\tilde{Y}}c_{3}(\tilde{Y}),
\end{equation}
with $a_{ij}=\frac{1}{2}\kappa_{iij}$ for $i\leq j$ and $a_{ij}=a_{ji}$. From now on, we work in the following gauge
\begin{equation}
X^0=1,
\qquad
X^i=z^i,
\qquad
\mathcal{F}_i=\partial_i\mathcal{F},
\qquad
\mathcal{F}_0=2\mathcal{F}-z^i\partial_i\mathcal{F},
\label{eq:general_period_vector}
\end{equation}
where $\partial_{i}\mathcal{F}=\partial\mathcal{F}/\partial z^{i}$. The period vector in \eqref{eq:general_periods} can then be written in the LCS regime as
\begin{equation}\label{eq:period-vector}
\mathbf{\Pi} = \begin{pmatrix} 
2\mathcal{F} - z^{i} \partial_i \mathcal{F} \\ 
\partial_i \mathcal{F}\\ 
1 \\ 
z^{i} 
\end{pmatrix} = \begin{pmatrix} 
\frac{1}{6}\kappa_{ijk}z^iz^j
z^k + b_iz^i - \frac{\chi\,\zeta(3)}{(2\pi i)^3} + \mathcal{O}(e^{2\pi i z}) \\
-\frac{1}{2}\kappa_{ijk}z^jz^k + a_{ij}z^j + b_i + \mathcal{O}(e^{2\pi i z})  \\ 
1 \\ 
z^{i}
\end{pmatrix}.
\end{equation}

\noindent The presence of closed string background fluxes generates a non-trivial scalar potential for the moduli fields in the EFT. Denoting by $H_{3}$ and $F_{3}$ the flux backgrounds corresponding to the 2-forms $B_{2}$ and $C_{2}$, one can generate a tree-level superpotential $W_{\text{flux}}$, known as the Gukov--Vafa--Witten (GVW) superpotential, for the $\mathcal{N}=1$ theory given by \cite{Gukov:1999ya}
\begin{equation}\label{eq:general_W0}
W_{\text{flux}}(z,\tau)=\int_{Y}(F_3-\tau H_3)\wedge\Omega(z)
=(\textbf{f}-\tau \textbf{h})^{T}\eta\hspace{0.5mm}\mathbf{\Pi}(z),\quad\eta=\begin{pmatrix}
    0&& \mathbb{I}\\
    -\mathbb{I}&& 0
\end{pmatrix},
\end{equation}
where $\mathbf{f}$ and $\mathbf{h}$ are the RR and NSNS integer flux vectors that originate from expanding the fluxes $F_{3}$ and $H_{3}$ in the symplectic basis $(\alpha_I,\beta^I)$ as 
\begin{equation}
    F_{3}=f_{2}^{I}\alpha_{I}-f_{1,I}\beta^{I},\quad H_{3}=h_{2}^{I}\alpha_{I}-h_{1,I}\beta^{I}.
\end{equation}
The modulus $\tau=C_0+i e^{-\phi}$ is the axio-dilaton and $\eta$ is a $2h^{2,1}+2$-dimensional symplectic matrix. To be totally precise, the surviving complex structure moduli under the orientifold projection, needed to reduce from $\mathcal{N}=2$ to $\mathcal{N}=1$, span a submanifold $\tilde{\mathcal{M}}_{\text{cs}}$ of $\mathcal{M}_{\text{cs}}$ of dimension $h^{2,1}_{-}$. Throughout this work, we will always assume orientifold projections with $h^{2,1}_{+}=0$, so that all complex structure moduli survive.

\noindent The NSNS and RR flux vectors split naturally into two parts depending on whether they couple to the periods $X^{I}$ or their duals $\mathcal{F}_{I}$. Specifically,
\begin{equation}
\mathbf{f}^{T}=(\mathbf{f_1}^{T},\mathbf{f_2}^{T}),
\qquad
\mathbf{h}^{T}=(\mathbf{h_1}^{T},\mathbf{h_2}^{T}),
\end{equation}
with components
\begin{equation}
\mathbf{f}^{T}_1=(f_{1,I}),
\quad
\mathbf{f}^{T}_2=(f_{2}^{I}),\quad \mathbf{h}^{T}_1=(h_{1,I}),\quad \mathbf{h}^{T}_2=(h_{2}^{I}),\quad I=0,\ldots,h^{2,1}.
\end{equation}
The lower sectors $\mathbf{f}_1,\mathbf{h}_1$ couple to the periods $X^I$, while the upper
sectors $\mathbf{f}_2,\mathbf{h}_2$ couple to the dual quantities $\mathcal{F}_I$. In terms of this decomposition, the superpotential \eqref{eq:general_W0} reads
\begin{equation}\label{eq:W0_split}
W_{\text{flux}}(z,\tau)
=
(f_{1,I}-\tau h_{1,I})X^I
-
(f_{2}^{I}-\tau h_{2}^{I})\mathcal{F}_I.
\end{equation}
For later convenience, we combine both flux vectors $\mathbf{f}$ and $\mathbf{h}$ into an enlarged $4(h^{2,1}+1)$-dimensional flux vector $\boldsymbol{\theta}$ and assemble the periods $(\mathcal{F}_{I},X^{I})$ together with the axio-dilaton $\tau$ in an enlarged period vector $\boldsymbol{\Xi}$. These read
\begin{equation}\label{eq:enlarged-period-vector}
\boldsymbol{\theta}= \begin{pmatrix} 
 
\mathbf{f}\\ 
 \mathbf{h}\\ 
\end{pmatrix},\qquad \boldsymbol{\Xi}= \begin{pmatrix} 
 
X^{I}\\ 
-\mathcal{F}_{I}\\
-\tau X^{I}\\
\tau \mathcal{F}_{I}\\
\end{pmatrix}.
\end{equation}
Based on these new vectors, the expression in \eqref{eq:W0_split} can be readily rewritten as
\begin{equation}\label{eq:W0_linear}
W_{\text{flux}}(z,\tau)=\boldsymbol{\theta}^{T}\cdot\boldsymbol{\Xi}(z,\tau)=\theta^{A}\Xi_{A},\qquad A=1,\ldots,4(h^{2,1}+1).
\end{equation}
Moreover, we point out that although the superpotential is protected against perturbative corrections \cite{Dine:1986vd}, it admits non-perturbative corrections from Euclidean D-branes wrapping internal cycles. These contributions will however not play a role in our later discussion.

\paragraph{Tadpole cancellation.} Fluxes are subject to a tadpole cancellation condition due to the fact that they are defined on a compact space $Y$. Charge conservation in this internal space requires a balance between the charges carried by the fluxes, O-planes and D-branes in the configuration. In O3/O7 orientifold models, there is a D7-brane tadpole where the negative RR O7-charge (in units of D7-charge) needs to be cancelled by adding a certain number of D7-branes. Additionally, there is a D3-brane tadpole which includes every possible source of $C_{4}$ fields, and can be written as follows
\begin{equation}\label{eq:tadpole-cancellation}
    2(N_{D3}-N_{\overline{D3}})+N_{\mathrm{flux}}=Q_{D3},
\end{equation}
where $N_{D3}/N_{\overline{D3}}$ stand for the number of spacetime-filling D$3$-branes/anti-branes in the configuration, with a factor of $2$ reflecting that we are working in the covering space. The D3-brane charge from the fluxes is defined as
\begin{equation}
N_{\mathrm{flux}}
=
\int_{Y} F_3\wedge H_{3}
=f_{1,I}h_{2}^{I}-h_{1,I}f_{2}^{I}=\mathbf{f}^{T}\eta\mathbf{h}.
\label{eq:tadpole}
\end{equation}
Finally, the RHS of the equation consists of all negative D3-brane charge contributions which arise from the presence of O3-planes and from the gravitational curvature of the divisors wrapped by both O7-planes and D7-branes. It turns out to be a topological term, in that it depends on the Euler number of the fixed orientifold locus,
\begin{equation}
    Q_{D3}=\frac{\chi_{f}}{2},\qquad \chi_{f}=\sum_{i}\chi(\Sigma_{i})+N_{O3}.
\end{equation}
with $N_{O3}$ denoting the number of O3-planes and $\Sigma_{i}$ the divisors wrapped by the O7-planes. In practice, the tadpole condition \eqref{eq:tadpole-cancellation} is often written as
\begin{equation}\label{eq:tadpole-bound}
N_{\mathrm{flux}}\leq L,
\end{equation}
where $L$ is the so-called tadpole bound. We will always consider the case where $L\leq Q_{D3}$. Note that, in theory, if we allow for a configuration with anti-branes, $L$ could go beyond $Q_{D3}$. However, it would never exceed a certain threshold, as a state with a certain number of anti D3-branes living on a flux background will eventually decay into a new state with only D3-branes and flux \cite{Kachru:2002gs, DeWolfe:2004qx, Denef:2004ze}. The tadpole condition in \eqref{eq:tadpole-bound} is crucial to guarantee a finite landscape of flux vacua \cite{Grimm:2023lrf, Bakker:2023xkt}.

\paragraph{Vacuum solutions.}
In order to find the vacuum configurations, one searches for minima of the F-term supergravity potential 
\begin{equation}\label{eq:potential-F-terms}
V_F=\frac{e^K}{\mathcal{V}^{2}}\left(K^{\alpha\bar\beta}D_\alpha W\,\overline{D_\beta W}-3\abs{W}^2\right),\quad D_{\alpha}W=\partial_{\alpha}W+(\partial_{\alpha}K)W,
\end{equation}
where $\mathcal{V}$ is the volume of $Y$ and $\alpha$ runs over all chiral superfields including complex structure moduli, Kähler moduli and axio-dilaton. The Kähler potential in \eqref{eq:potential-F-terms} contains two pieces, one carrying the axio-dilaton $\tau$ and a second one, $K_{\text{cs}}$, dependent on the $z^{i}$ moduli through the period vector $\mathbf{\Pi}(z)$. They take the known form
\begin{equation}\label{eq:Kahler-potential}
    K=-\log{(-i(\tau-\bar{\tau}}))-\log\left(-i\bar{\mathbf{\Pi}}^{T}\eta\mathbf{\Pi}\right).
\end{equation}
Controlled vacua arise when stabilising the complex structure moduli and the axio-dilaton to a supersymmetric minimum, described by the vanishing of the covariant derivatives
\begin{equation}\label{eq:SUSY-vacua-equations}
    D_{\tau}W=0,\quad D_{i}W=0,\quad i=1,\ldots h^{2,1}.
\end{equation}
Here $D_{i}\equiv D_{z^{i}}$. Solutions to these equations are called imaginary self-dual (ISD) vacua, since the conditions \eqref{eq:SUSY-vacua-equations} can be shown to be equivalent to the imaginary self-duality constraint $*G_{3}=iG_{3}$, where $G_{3}=F_{3}-\tau H_{3}$ and $*$ is the Hodge star in $Y$ \cite{Giddings:2001yu}. In fact, there is one more way to write the vacuum solutions in terms of the gauge kinetic matrix $\mathcal{N}_{IJ}$ of the underlying $\mathcal{N}=2$ geometry. They take the form 
\begin{equation}\label{eq:third-way-ISD}
    \mathbf{f}_{1}-\tau\mathbf{h}_{1}=\bar{\mathcal{N}}\cdot(\mathbf{f}_{2}-\tau\mathbf{h}_{2}),
\end{equation}
where $\mathcal{N}_{IJ}$ is the gauge kinetic matrix defined as
\begin{equation}
    \mathcal{N}_{IJ}=\bar{\mathcal{F}}_{IJ}+2i\frac{\text{Im}(\mathcal{F}_{IK})X^{K}\text{Im}(\mathcal{F}_{JL})X^{L}}{X^{M}\text{Im}(\mathcal{F}_{MN})X^{N}}, \qquad \mathcal{F}_{IJ}=\partial_{I}\partial_{J}\mathcal{F},
\end{equation}
with $\partial_{I}=\partial/\partial X^{I}$. Combining the components $\mathbf{f}_{i}$ and $\mathbf{h}_{i}$ into $\mathbf{f}$ and $\mathbf{h}$, one can turn the condition \eqref{eq:third-way-ISD} into
\begin{equation}\label{eq:vacua-condition}
    \mathbf{f}=(\tau_{1}\mathbb{I}+\tau_{2}\eta\mathcal{M})\mathbf{h},
\end{equation}
where we have introduced the so-called ISD matrix $\mathcal{M}$, whose entries depend on the real and imaginary parts of $\mathcal{N}_{IJ}$, denoted respectively by $\mathcal{R}$ and $\mathcal{I}$. Explicitly,
\begin{equation}\label{eq:ISD-matrix}
    \mathcal{M}=\begin{pmatrix}
    -\mathcal{I}^{-1}&& \mathcal{I}^{-1}\mathcal{R}\\
    \mathcal{R}\mathcal{I}^{-1}&& -\mathcal{I}-\mathcal{R}\mathcal{I}^{-1}\mathcal{R}
\end{pmatrix}.
\end{equation}
Studying the components of this matrix will be of central interest in section \ref{sec:fine-tuning-and-distance}.

Since we are just going to work with the flux superpotential, we set $W=W_{\text{flux}}$ from now on. In fact, it is instructive to deal with the vacuum expectation value (VEV) of its gauge invariant\footnote{Recall that under Kähler transformations, $K\rightarrow K+f+\bar{f}$ and $W\rightarrow e^{-f}W$, so that $e^{K}\abs{W}^2$ remains invariant. This is consistent with the fact that $e^{K}\abs{W}^2$ determines the gravitino mass, which is a physical quantity that cannot depend on the gauge.} version, which is defined as
\begin{equation}\label{eq:gauge-inv-superpotential}
    W_{0}=\sqrt{\frac{2}{\pi}}\left\langle e^{K/2}W\right\rangle.
\end{equation}
Remarkably, once stabilised, the Kähler moduli values are proportional to $\log\abs{W_{0}}^{-1}$, so control over the EFT expansion in $\alpha^{\prime}$ demands small values of $\abs{W_{0}}$, which exist for a highly fine-tuned selection of fluxes. This motivates the goal of quantifying the degree of fine-tuning required to achieve such vacua using Fisher information techniques, which we review in section \ref{sec:fisher}. Before turning to this, let us set the stage by introducing a particular CY compactification that we will use as our main example later on.

\subsection{Specialisation to \texorpdfstring{$\mathbb{P}^4(1,1,1,6,9)$}{P4[1,1,1,6,9]} at large complex structure}\label{sec:specialisation-geoemtry}

To make the general discussion more specific, let us stick to a particular example, which is  the one discussed in \cite{Chauhan:2025rdj}. We consider Type IIB compactified on a $\mathbb{Z}_{2}$ involution\footnote{This involution has been discussed in \cite{Louis:2012nb}. We have two O7-planes and eight D7-branes (four on each O-plane), so that there is no D7-brane tadpole. For this model, one has $Q_{D3}=138$.} of the degree 18 hypersurface in the weighted projective space $\mathbb{P}^4(1,1,1,6,9)$, with corresponding Hodge numbers $h^{1,1}=2$ and $h^{2,1}=272$, studied in \cite{Candelas:1994hw}. 

By restricting the analysis to the symmetric locus of its $\mathbb{Z}_{6}\times\mathbb{Z}_{18}$ discrete symmetry, the complex structure moduli invariant under this symmetry reduce to two $h^{2,1}=2$. It is worth noting that the $\mathbb{Z}_{2}$ involution in the corresponding orientifold quotient keeps both complex structure moduli, i.e. $h^{2,1}_{-}=2$. These two effective moduli are immediately realised as the deformation parameters $\psi$ and $\phi$ of the corresponding hypersurface parametrised by
\begin{equation}
    x_{1}^{18}+x_{2}^{18}+x_{3}^{18}+x_{4}^{3}+x_{5}^2-18\psi x_{1}x_{2}x_{3}x_{4}x_{5}-3\phi x_{1}^6 x_{2}^6 x_{3}^6=0.
\end{equation}
In order to specify the prepotential \eqref{eq:general_prepotential} underlying this geometry, we use the topological data of the mirror $X_{18}(1,1,1,6,9)$. The non-zero intersection numbers are
\begin{equation}\label{eq:kappas_model}
\kappa_{111}=9,\quad
\kappa_{112}=3,\quad\kappa_{122}=1,
\end{equation}
and the remaining topological constants are
\begin{equation}
a_{11}=\frac92,\qquad
a_{12}=a_{21}=\frac32,\qquad
b_1=\frac{17}{4},\qquad
b_2=\frac{3}{2}.
\label{eq:bij_ai_model}
\end{equation}
Having established the geometric setup and specified our benchmark model, we now turn to the information theoretic framework that will be used to quantify fine-tuning across the flux landscape.

\section{Fisher Information in the Flux Vacua Landscape}\label{sec:fisher}

As already mentioned in the introduction, the main idea of the work \cite{Halverson:2026cxe} was to develop a new measure of naturalness or fine-tuning from an information-theoretic rather than a geometric perspective. We briefly summarise now some useful notions that will be relevant for us later on.

The question to keep in mind is: how does a set of parameters influence a set of observables? An intuitive possibility is to think of the set of parameters $\lbrace\theta^{A}\rbrace_{A=1,\ldots,N}$ as the UV couplings of some physical theory that influence a set of observables $\lbrace x^{a}\rbrace_{a=1,\ldots,n}$ in the IR after following the RG trajectories. The idea is to assign to every point $\theta$ in parameter space a theoretical probability distribution $p_{\theta}(x)$ over the observable space, centred around a point $X(\theta)$, where $X$ represents the map from parameter space to the predictions in observable space. Clearly,  if one has sufficient predictability on the resulting IR observables, a natural option is to consider $p_{\theta}(x)=\delta^{(n)}(x^{a}-X^{a}(\theta))$, simply localised on the predicted value. However, this is not usually the case, and one prefers to soften the sharp behaviour of the delta function by introducing a width $\sigma$ acting as a regulator. A standard example is the Gaussian profile\footnote{In general, any distribution that introduces a regulator may be of use. A second choice mentioned in \cite{Halverson:2026cxe} is that of a uniform distribution, where $p_{\theta}(x)$ is constant inside a ball of radius $\epsilon$ (regulator) and zero elsewhere.}
\begin{equation}\label{eq:gaussian-profiles}
    p_{\theta}(x)=\prod_{a=1}^{n}\frac{1}{\sqrt{2\pi}\sigma}\text{ exp}\left(-\frac{(x^{a}-X^{a}(\theta))^2}{2\sigma^{2}}\right).
\end{equation}
In this setting, the notion of fine-tuning or, equivalently, naturalness reduces to studying the sensitivity of $p_{\theta}$ to small perturbations in parameter space of the form $\theta+\delta\theta$. To this end, one introduces the notion of $\textit{divergence}$, a formal definition of “distance” between two probability distributions, to which we now turn.

\subsection{From divergences to a pullback information metric}

A standard way to endow a parameter space \(\mathcal{P}\) with a local
information geometry is to start from a so-called divergence \(d\) between two probability
densities \(p_{\theta}(x)\) and \(p_{\theta'}(x)\), where
\(\theta,\theta'\in\mathcal{P}\).
Following the general discussion in \cite{Halverson:2026cxe}, a divergence is defined as a
map \cite{amari2000methods, ay2017information}
\begin{equation}\label{eq:divergence}
d:\ (p_{\theta},p_{\theta'})\mapsto d(p_{\theta},p_{\theta'})
\in \mathbb{R}_{\ge 0}
\end{equation}
such that
\begin{equation}
d(p_{\theta},p_{\theta'})\ge 0,
\qquad
d(p_{\theta},p_{\theta'})=0
\iff
p_{\theta}=p_{\theta'},
\end{equation}
and whose expansion for small displacements defines a positive semi-definite local quadratic form,
\begin{equation}\label{eq:distance-expanded}
d(p_{\theta},p_{\theta+\delta\theta})
=
\frac12\,g_{AB}(\theta)\,\delta\theta^A\delta\theta^B
+\mathcal{O}(\delta\theta^3).
\end{equation}
In general, \(d\) need not be symmetric or satisfy the triangle
inequality, so it is not necessarily a metric in the strict distance-theoretic
sense. One may alternatively express the divergence $d$ above in terms of a diagonal basis of eigenvectors $\delta\varphi^{A}$ as follows
\begin{equation}\label{eq:distance-diagonalised}
    d(p_{\varphi},p_{\varphi+\delta\varphi})=\frac{1}{2}\lambda_{A}(\delta\varphi^{A})^2+\mathcal{O}(\delta\varphi^3),
\end{equation}
where $\lambda_{A}$ denotes the $g_{AB}$ eigenvalue associated with $\delta\varphi^{A}$, and summation over $A$ is implied. This second representation of $d$ gives us a very intuitive picture of the naturalness measure as we discuss below.

Additionally, one typically further requires the metric $g_{AB}$ to be invariant under sufficient statistics\footnote{By sufficiency here we mean that all $\theta$-dependence in $p_{\theta}$ goes through the sufficient statistic $S(x)$, so that $p_{\theta}(x)=\tilde{p}_{\theta}(S(x))\hspace{0.5mm}\hat{p}(x)$, where $\hat{p}$ is a $\theta$-independent function.} for $\theta$. Stated more concretely, one does not expect the measure of fine-tuning to depend on the  sufficient statistics, which are realised as subsets of observables encoding all the dependence on the parameters. Remarkably, this extra condition fixes the metric, up to an overall constant, to be the Fisher Information Metric (FIM) \cite{cencov2000statistical, ay2015information}
\begin{equation}\label{eq:authentic-FIM}
    I_{AB}(\theta)=\int d^{n}x\hspace{1mm}p_{\theta}(x)\frac{\partial\log{p_{\theta}(x)}}{\partial\theta^{A}}\frac{\partial\log{p_{\theta}(x)}}{\partial\theta^{B}}.
\end{equation}
Roughly, this metric encodes the amount of information carried by the distributions $p_{\theta}(x)$ about the parameters $\theta$. For the case of Gaussian profiles \eqref{eq:gaussian-profiles} and after removing the dependence on the corresponding regulator\footnote{We do not want the final metric to depend on the way we have regulated the sharpness of the delta function.} $\sigma$, the FIM in \eqref{eq:authentic-FIM} reads
\begin{equation}\label{eq:FIM}
    F_{AB}(\theta)=\frac{\partial X^{a}}{\partial\theta^{A}}\frac{\partial X^{b}}{\partial\theta^{B}}\delta_{ab}.
\end{equation}
Notably, when $n\geq N$, this metric admits a clear geometrical interpretation as the pullback of the flat observable space metric onto the $N$-dimensional hypersurface traced by the admissible model observations. By studying the eigenvalues of $F_{AB}$ we can learn about which directions in parameter space are relevant or irrelevant for a change in the observables. Namely, glancing at \eqref{eq:distance-diagonalised}, we see that small eigenvalues would be associated with directions in $\mathcal{P}$ along which the observables remain virtually unchanged, whereas large eigenvalues would correspond to directions in $\mathcal{P}$ for which the observable probability densities exhibit bigger changes. Small changes in $\mathcal{P}$ implying small (big) changes in the observable space are called \textit{natural} (\textit{unnatural}). Fine-tuning refers precisely to the presence of such unnatural directions, as in order to reproduce a concrete value for the observables, one has to restrict to a very specific (fine-tuned) configuration in parameter space.

Before turning to the application, let us present another way to write the FIM in terms of the $n\times N$ Jacobian matrix $J$ of the admissible model observations $X^{a}(\theta)$, which can be expressed as
\begin{equation}\label{eq:Jacobian}
    J = \begin{pmatrix}
\frac{\partial X^1}{\partial \theta^1} & \frac{\partial X^1}{\partial \theta^2} & \cdots & \frac{\partial X^1}{\partial \theta^N} \\[0.3em]
\frac{\partial X^2}{\partial \theta^1} & \frac{\partial X^2}{\partial \theta^2} & \cdots & \frac{\partial X^2}{\partial \theta^N} \\
\vdots & \vdots &  & \vdots \\
\frac{\partial X^n}{\partial \theta^1} & \frac{\partial X^n}{\partial \theta^2} & \cdots & \frac{\partial X^n}{\partial \theta^N}
\end{pmatrix}.
\end{equation}
Based on it, the FIM in \eqref{eq:FIM} can be compactly expressed as
\begin{equation}
    F=J^{T}J.
\end{equation}
Regarding the zero eigenvalues of the Fisher matrix, it is easy to check that the null spaces (or kernels) of $F$ and $J$ are identical. Indeed, a vector $v\in\text{Ker}(F)$ satisfies
\begin{equation}
    Fv=J^{T}Jv=0.
\end{equation}
Contracting both sides with $v^{T}$ we find that
\begin{equation}
    v^{T}J^{T}Jv=(Jv)^T Jv=\norm{Jv}^2=0,
\end{equation}
which holds if and only if $Jv=0$ and thus $v\in\text{Ker}(J)$. Conversely, for a given $v\in\text{Ker}(J)$, one straightforwardly has
\begin{equation}
    J^{T}Jv=Fv=0,
\end{equation}
and so $v\in\text{Ker}(F)$. Hence, we conclude that Ker$(F)=\text{Ker}(J)$. Now, the rank-nullity theorem applied to the Jacobian map tells us that
\begin{equation}
    \text{dim(Ker}(J))+\text{Rank}(J)=N,
\end{equation}
with $\text{Rank}(J)$ being the dimension of the image of $J$. Using this constraint together with the identification of kernels, we arrive at
\begin{equation}
    \text{dim(Ker}(F))=\text{dim(Ker}(J))=N-\text{Rank}(J)\geq N-n,
\end{equation}
where the final inequality follows from Rank$(J)\leq n$, as $J$ is an $n \times N$ matrix and we are assuming that $n\leq N$. Since we will be interested in just one observable $X$ ($n=1$), the FIM will have $N-1$ zero eigenvalues and a unique non-trivial positive eigenvalue from
\begin{equation}\label{eq:eigenvalue-eq-jacobian}
    FJ^{T}=J^{T}(J J^{T})=\norm{J}^2 J^{T}.
\end{equation}
The Euclidean norm $\norm{J}^2$ is positive and its corresponding eigenvector is just the gradient vector $J^{T}=\nabla_{\theta}X=(\partial_{\theta^{1}}X,\ldots,\partial_{\theta^{N}}X)^T$. It is precisely this eigenvalue $\norm{\nabla_{\theta}X}^2$ that will measure the amount of fine-tuning of our observable $X$ along the direction dictated by $\nabla_{\theta}X$. We close by noting that, as already pointed out in \cite{Halverson:2026cxe}, the measure of fine-tuning $\norm{\nabla_{\theta}X}^2$ looks remarkably similar to previous proposals, such as the logarithmic version in \cite{barbieri1988upper}
\begin{equation}
    \rho(\theta)=\left(\frac{\partial\hspace{0.5mm}\text{log}\hspace{0.5mm}X(\theta)}{\partial\hspace{0.5mm}\text{log}\hspace{0.5mm}\theta}\right)^2,
\end{equation}
where in this case we are simply considering one parameter $\theta$. Thus, the FIM naturally reduces to this previously studied case and in some sense upgrades the single-parameter measure by accounting for correlations of the $\theta$ parameters encoded in the off-diagonal terms of $F_{AB}$.

\subsection{Application to the flux landscape}

We now apply the Fisher Information techniques just introduced to flux vacua in the landscape. As already mentioned towards the end of section \ref{subsec:periods}, our goal is to quantify the amount of fine-tuning required to reproduce a well-controlled vacuum solution. We are going to work with the ``physical" or gauge invariant version of the superpotential, introduced in \eqref{eq:gauge-inv-superpotential}. In particular, the observable of interest is given by its modulus
\begin{equation}\label{eq:Xobservable}
X(\theta)=\abs{W_0(\theta,\tau(\theta),z(\theta))}.
\end{equation}
Since we are interested in vacuum solutions, it is clear that $\tau$ and $z^{i}$ will depend on the fluxes, a dependence that can be determined by solving the equations \eqref{eq:SUSY-vacua-equations}. Thus, all the freedom comes from tweaking the value of the fluxes, which span the parameter space $\mathcal{P}$. For only one observable, the FIM \eqref{eq:FIM} simply reduces to
\begin{equation}\label{eq:FIM-superpotential}
F_{AB}(\theta)
=\frac{dX}{d\theta^{A}}\frac{dX}{d\theta^{B}}.
\end{equation}
Here, the total derivatives account for both the explicit $\theta$-dependence and the implicit dependence entering through the stabilised moduli. The metric \eqref{eq:FIM-superpotential} is a $4(h^{2,1}+1)$-dimensional matrix whose eigenvectors tell us about the specific directions in the flux lattice that drive changes in the superpotential. A large eigenvalue corresponds to a highly sensitive direction, meaning that just a small change in the fluxes along the associated direction could take us away from a carefully tuned small $\abs{W_{0}}$ vacuum, whereas a small eigenvalue corresponds to an almost irrelevant deformation in the flux space that keeps the observable within a desired small superpotential region.

Let us comment on the fact that the fluxes are quantised, $\boldsymbol{\theta}\in\mathbb{Z}^{4(h^{2,1}+1)}$. The divergence between the distributions associated with two flux configurations is well defined for discrete $\boldsymbol{\theta}$. The Fisher metric, however, arises from its second-order expansion around $\theta^{A}$ and therefore requires differentiability. In the Fisher analysis we thus treat the fluxes as continuous variables, working with the smooth extension of $|W_0|$ to real fluxes. Treating the flux lattice as a continuum is also standard in the statistical analysis of flux vacua, where sums over integral fluxes are approximated by integrals \cite{Ashok:2003gk,Denef:2004ze}. As there, this approximation is reliable when a unit change in the fluxes is small compared with the flux vector itself, a condition limited here by the tadpole bound. The Fisher eigenvalue should therefore be understood as capturing the response of $|W_0|$ to small flux deformations, keeping in mind that the physical fluxes are integers. As we will see in section \ref{sec:fine-tuning-and-distance}, deep in the large complex structure regime at weak coupling this discreteness has non-trivial consequences, as flux integrality, combined with the tadpole bound, forces some flux components to vanish.

\begin{figure}[h]
    \centering
    \includegraphics[width=0.6\linewidth]{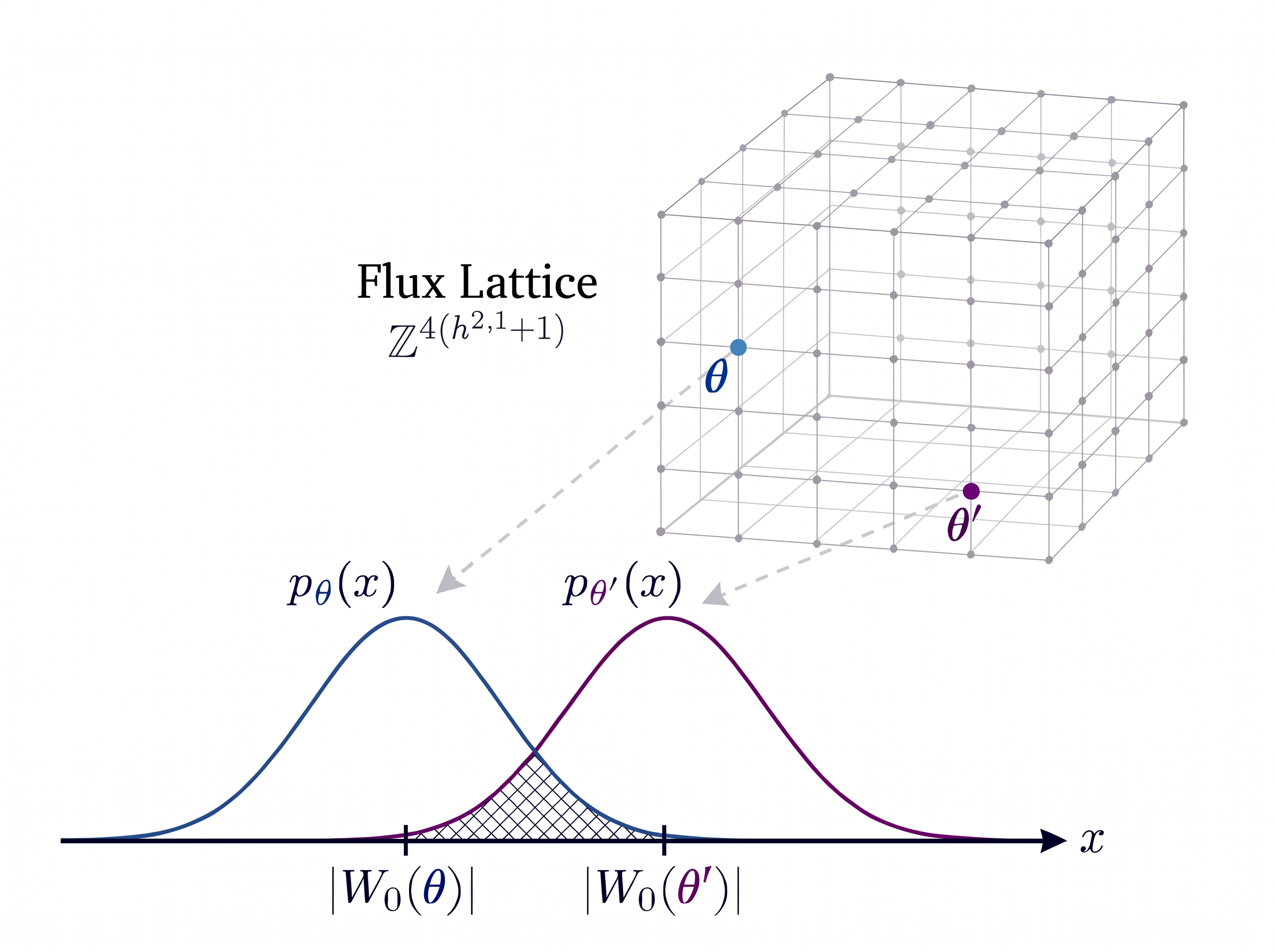}
    \caption{Sketch showing the parameter space as the multi-dimensional flux lattice along with the observable space defined as the real line where the quantity $\abs{W_0}$ takes values. Any point $\theta$ in the flux space has a corresponding Gaussian distribution in observable space centred around $\abs{W_{0}(\theta)}$, given by $p_{\theta}(x)=\frac{1}{\sqrt{2\pi}\sigma}\text{ exp}\left(-\frac{(x-\abs{W_{0}(\theta)})^2}{2\sigma^{2}}\right)$. The degree of overlap between the two Gaussians is measured by the divergence $d$ introduced in \eqref{eq:divergence}.}
    \label{fig:fisher-superpotential}
\end{figure}

\noindent Based on \eqref{eq:Xobservable}, the total derivative with respect to the fluxes is computed by means of the chain rule, giving
\begin{equation}\label{eq:chain-rule}
    \frac{d\abs{W_{0}}}{d\theta^{A}}=\frac{\partial\abs{W_{0}}}{\partial\theta^{A}}+\frac{\partial\abs{W_{0}}}{\partial\tau}\frac{\partial\tau}{\partial\theta^{A}}+\frac{\partial\abs{W_{0}}}{\partial\bar{\tau}}\frac{\partial\bar{\tau}}{\partial\theta^{A}}+\frac{\partial\abs{W_{0}}}{\partial z^{i}}\frac{\partial z^{i}}{\partial\theta^{A}}+\frac{\partial\abs{W_{0}}}{\partial \bar{z}^{i}}\frac{\partial \bar{z}^{i}}{\partial\theta^{A}}.
\end{equation}
Using the definition \eqref{eq:gauge-inv-superpotential}, one computes the $\tau$-derivative
\begin{equation}
    \partial_{\tau}\abs{W_{0}}=\sqrt{\frac{2}{\pi}}e^{K/2}\left(\partial_{\tau}\abs{W}+\frac{1}{2}\abs{W}\partial_{\tau}K\right),
\end{equation}
which reduces to
\begin{equation}\label{eq:derivative-tau-W0}
    \partial_{\tau}\abs{W_{0}}=\sqrt{\frac{2}{\pi}}e^{K/2}\frac{\bar{W}}{2\abs{W}}\left(\partial_{\tau}W+W\hspace{0.5mm}\partial_{\tau}K\right)=\sqrt{\frac{2}{\pi}}e^{K/2}\frac{\bar{W}}{2\abs{W}}D_{\tau}W\hspace{0.5mm},
\end{equation}
where we have used that
\begin{equation}
    \partial_{\tau}\abs{W}=\frac{\bar{W}}{2\abs{W}}\partial_{\tau}W\hspace{0.5mm},
\end{equation}
which follows from the holomorphicity of $W$. Similarly, for the complex structure moduli piece in \eqref{eq:chain-rule}, one obtains the same structure as in \eqref{eq:derivative-tau-W0} again by holomorphicity. That is,
\begin{equation}\label{eq:derivative-cs-W0}
    \partial_{i}\abs{W_{0}}=\sqrt{\frac{2}{\pi}}e^{K/2}\frac{\bar{W}}{2\abs{W}}D_{i}W\hspace{0.5mm}.
\end{equation}
When evaluated at a vacuum solution, both \eqref{eq:derivative-tau-W0} and \eqref{eq:derivative-cs-W0} vanish according to \eqref{eq:SUSY-vacua-equations}. Hence, the intricate dependence of $\tau$ and $z^{i}$ on the flux parameters obtained by solving the coupled system of algebraic equations \eqref{eq:SUSY-vacua-equations} and needed to compute the derivatives $\partial_{A}z^{i},\partial_{A}\bar{z}^{i},\partial_{A}\tau, \partial_{A}\bar{\tau}$ in \eqref{eq:chain-rule} drops out of the Fisher matrix evaluated on a vacuum solution. Recalling \eqref{eq:W0_linear}, the explicit dependence of $\abs{W_{0}}$ on $\theta$ is captured by
\begin{equation}\label{eq:components-eigenvector}
    \frac{\partial\abs{W_{0}}}{\partial\theta^{A}}=\frac{e^{K/2}}{\sqrt{2\pi}\abs{W}}\left(\Xi_{A}\bar{W}+\bar{\Xi}_{A}W\right)=\sqrt{\frac{2}{\pi}}\frac{e^{K/2}}{\abs{W}}\text{Re}(\Xi_{A}\bar{W}).
\end{equation}
This allows us to write the FIM in \eqref{eq:FIM-superpotential} as
\begin{equation}\label{eq:final-FIM}
    F_{AB}(\theta)=\frac{2e^{K}}{\pi\abs{W}^2}\text{Re}(\Xi_{A}\bar{W})\hspace{0.5mm}\text{Re}(\Xi_{B}\bar{W}),
\end{equation}
or alternatively,
\begin{equation}\label{eq:fisher-matrix-argument-superpotential}
    F_{AB}(\theta)=\frac{2e^{K}}{\pi}\text{Re}(e^{-i\varphi}\hspace{0.5mm}\Xi_{A})\hspace{0.5mm}\text{Re}(e^{-i\varphi}\hspace{0.5mm}\Xi_{B}),
\end{equation}
where we have introduced the phase $\varphi$ of the flux superpotential $W=\abs{W}\hspace{0.5mm}e^{i\varphi}$. 
As already mentioned in the previous section, this FIM will have only one non-trivial eigenvalue $\lambda=\norm{\nabla_{\theta}\abs{W_{0}}}^2$, which can of course be realised as the trace of the Fisher matrix in \eqref{eq:FIM-superpotential}. Let us proceed by analysing more deeply the dependence of this eigenvalue on the moduli space coordinates. 

\subsection{An exact factorisation of $\lambda$}
The eigenvalue $\lambda$ of the Fisher matrix $F_{AB}$ admits a clear factorisation into one piece which depends only on the axio-dilaton $\tau$, a second piece bringing only complex structure moduli dependence and a third piece which depends both on $\tau$ and $z^{i}$ but can be bounded and therefore remains under control. The starting point here is just the definition of $\lambda$ as the trace of the matrix \eqref{eq:fisher-matrix-argument-superpotential},
\begin{equation}\label{eq:definition-trace}
    \lambda=\text{Tr}(F_{AB})=\frac{2e^{K}}{\pi}\sum_{A=1}^{N}\left(\text{Re}(e^{-i\varphi}\hspace{0.5mm}\Xi_{A})\right)^2.
\end{equation}
To shorten notation, let us define the phase $u=e^{-i\varphi}$. The sum over the flux parameters gives rise to two terms according to the decomposition
\begin{equation}\label{eq:expansion-real-part}
    \sum_{A=1}^{N}\left(\text{Re}(u\hspace{0.5mm}\Xi_{A})\right)^2=\frac{1}{4}\sum_{A=1}^{N}(u^2\Xi_{A}^2+\bar{u}^2\bar{\Xi}_{A}^2+2\abs{\Xi_{A}}^2)=\frac{1}{2}\left(\norm{\boldsymbol{\Xi}}^2+\text{Re}(u^2\hspace{0.5mm}\boldsymbol{\Xi}^{T} \boldsymbol{\Xi})\right).
\end{equation}
Here, we simply used that $\abs{u}=1$ and introduced the standard Hermitian squared norm, which is defined as
\begin{equation}\label{eq:hermitian-norm}
\norm{\boldsymbol{\Xi}}^2=\boldsymbol{\Xi}^{\dagger}\boldsymbol{\Xi}=\sum_{A=1}^{N}\abs{\Xi_{A}}^2.
\end{equation}
In view of \eqref{eq:expansion-real-part}, $\lambda$ takes the form
\begin{equation}\label{eq:lambda-mid-decomposition}
    \lambda=\frac{e^{K}}{\pi}\norm{\boldsymbol{\Xi}}^2\left(1+\frac{\text{Re}(u^2\hspace{0.5mm}\boldsymbol{\Xi}^{T}\boldsymbol{\Xi})}{\norm{\boldsymbol{\Xi}}^2}\right).
\end{equation}
We first analyse the factor outside the brackets. By definition, the Kähler potential in \eqref{eq:Kahler-potential} factorises into two terms for $\tau$ and $z^{i}$ respectively, yielding $e^{K}=e^{K_{\text{cs}}}/(2\hspace{0.5mm}\tau_{2})$, where $\tau_{2}=\text{Im}\tau$. Similarly, the norm of the enlarged period vector defined in \eqref{eq:enlarged-period-vector} decomposes into
\begin{equation}
    \norm{\boldsymbol{\Xi}}^2=\norm{\eta\mathbf{\Pi}}^2+\norm{-\tau\eta\mathbf{\Pi}}^2=(1+\abs{\tau}^2)\norm{\mathbf{\Pi}}^2,
\end{equation}
since the orthogonality of $\eta$ yields $\norm{\eta\mathbf{\Pi}}^2=\mathbf{\Pi}^{\dagger}\eta^{T}\eta\mathbf{\Pi}=\mathbf{\Pi}^{\dagger}\mathbf{\Pi}=\norm{\mathbf{\Pi}}^2$. In summary, $\lambda$ in \eqref{eq:lambda-mid-decomposition} can be recast as a product of three factors
\begin{equation}\label{eq:lambda-decomposition}
    \lambda=\frac{2}{\pi}\mathcal{A}\hspace{0.6mm}D(\tau)\hspace{0.6mm}G(z),
\end{equation}
where we have defined the functions
\begin{equation}\label{eq:functions-composing-lambda}
    D(\tau)=\frac{1+\abs{\tau}^2}{\tau_{2}},\qquad G(z)=\frac{1}{4}e^{K_{\text{cs}}}\norm{\mathbf{\Pi}}^2,\qquad \mathcal{A}=1+\frac{\text{Re}(u^2\hspace{0.5mm}\boldsymbol{\Xi}^{T}\boldsymbol{\Xi})}{\norm{\boldsymbol{\Xi}}^2}.
\end{equation}
Note that unlike $D(\tau)$ and $G(z)$, the function $\mathcal{A}$ does depend on both $\tau$ and $z^{i}$. However, it is not hard to check that it is bounded both from below and above. This can be seen by bounding the numerator
\begin{equation}
    \lvert\text{Re}(u^2\hspace{0.5mm}\boldsymbol{\Xi}^{T}\boldsymbol{\Xi})\rvert\leq\abs{u}^2\hspace{0.5mm}\lvert\boldsymbol{\Xi}^{T}\boldsymbol{\Xi}\rvert=\lvert\boldsymbol{\Xi}^{T}\boldsymbol{\Xi}\rvert,
\end{equation}
which can be further linked to the denominator by means of Cauchy-Schwarz's inequality
\begin{equation}\label{eq:cauchy-inequality}
\lvert{\boldsymbol{\Xi}^{T}\boldsymbol{\Xi}}\rvert\leq\sum_{A=1}^{N}\abs{\Xi_{A}}^2=\norm{\boldsymbol{\Xi}}^2.
\end{equation}
Finally, defining the parameter $\varepsilon$ as
\begin{equation}
    \varepsilon=\frac{\lvert\mathbf{\Xi}^{T}\mathbf{\Xi}\rvert}{\norm{\mathbf{\Xi}}^2}\in[0,1],
\end{equation}
it follows that
\begin{equation}\label{A-vs-epsilon-region}
    1-\varepsilon\leq\mathcal{A}\leq 1+\varepsilon.
\end{equation}
That is to say, $\mathcal{A}$ lies in a band whose width is set by $\varepsilon$, and thus always takes values in the interval $\mathcal{A}\in[0,2]$. We will see later that along certain asymptotic regions in moduli space, the value of $\mathcal{A}$ tends to saturate the upper bound.

Before moving on, let us make two important remarks. Firstly, $\lambda$ does not depend explicitly on the fluxes. As follows from \eqref{eq:components-eigenvector}, the gradient $\partial_A|W_0|$ depends only on the values of the moduli at the vacuum and on the phase of $W$. Two vacua with the same moduli and the same phase therefore have the same $\lambda$, whatever their fluxes. In particular, since the flux superpotential \eqref{eq:W0_linear} is linear in $\theta^A$, the vacuum conditions are invariant under a rescaling $\theta^{A}\to t\,\theta^{A}$. The moduli are then unchanged, $|W_0|$ is homogeneous of degree one, and its gradient (and hence $\lambda$) is homogeneous of degree zero.

Secondly, $\lambda$ depends on the duality frame. Vacua related by $SL(2,\mathbb{Z})$ or $Sp(2h^{2,1}+2,\mathbb{Z})$ transformations are physically equivalent but are in general assigned different values of $\lambda$. These transformations act linearly on the fluxes, $\theta^{A} \to \tensor{\Lambda}{^{A}_{B}}\,\theta^{B}$, together with a corresponding transformation of the moduli, and leave $|W_0|$ invariant. Its gradient therefore transforms covariantly, $\nabla_\theta|W_0|\to(\Lambda^{T})^{-1}\nabla_\theta|W_0|$, but its Euclidean norm $\lambda$ is preserved only if $\Lambda$ is orthogonal, which is not the case for these integral symplectic matrices. The frame
dependence of $\lambda$ thus originates from the Euclidean metric implicitly used on the space of fluxes, and not from the observable itself. It is therefore crucial to fix a duality frame. Throughout this work we use, as is customary, the integral symplectic basis adapted to the large complex structure point, in which the periods take the form \eqref{eq:period-vector}, and we restrict the moduli to a fundamental domain: the axio-dilaton takes values in the fundamental domain of $SL(2,\mathbb{Z})$, which we denote by $\mathcal{M}_\tau$,
\begin{equation}
  \mathcal{M}_\tau=\Big\{\tau_1\in\big(-\tfrac12,\tfrac12\big],\ |\tau|>1\Big\}
  \cup\Big\{|\tau|=1,\ \tau_1\in\big[0,\tfrac12\big]\Big\},\quad \tau_{1}=\text{Re}\hspace{0.5mm}\tau,
\end{equation}
and the axions take values $x^i=\operatorname{Re}z^i\in(-\tfrac12,\tfrac12]$. We refer to this choice as the canonical frame.

Fixing the frame in this way amounts to a gauge fixing of the duality group. A vacuum found with moduli outside this domain can always be mapped into it by a duality transformation, acting simultaneously on the moduli and on the fluxes, and is then identified with a vacuum already
described in the canonical frame. Each physically distinct vacuum thus has a unique representative, and $\lambda$ becomes a well-defined function on the space of vacua. This dependence on the parametrisation is inherent to the measure of \cite{Halverson:2026cxe}, which, as noted there, is not reparametrisation invariant: naturalness is a property of a model together with a prior on its parameters. Indeed, while the Fisher matrix is a tensor on parameter space, its eigenvalues rely on a metric on that space, here the Euclidean metric on the flux quanta in the chosen integral basis.

\subsection{Some useful formulas and a bound for $\abs{W_{0}}$}\label{sec:an-exact-formula-for-W0}

In this subsection, we derive some identities that will be needed in the next section and an upper bound for $\abs{W_{0}}$ in terms of the Fisher eigenvalue. To set the stage, we begin by defining the Hodge norm of a real $3$-form $v\in H^{3}(Y,\mathbb{R})$ as
\begin{equation}\label{eq:hodge-norm}
\norm{v}_{\mathcal{M}}^2=\int_{Y}v\wedge*v=2e^{K_{\text{cs}}}\left(\abs{\mathbf{v}^{T}\eta\mathbf{\Pi}}^2+G^{i\bar{j}}(\mathbf{v}^{T}\eta D_{i}\mathbf{\Pi})\overline{(\mathbf{v}^{T}\eta D_{j}\mathbf{\Pi})}\right).
\end{equation}
A few comments are in order. The vector $\mathbf{v}$ encodes the components of $v$ when expanded in the symplectic real basis $(\alpha_{I},\beta^{J})$. Regarding the expansion of $v$ in the complex basis $\lbrace\Omega,D_{i}\Omega,\overline{D_{i}\Omega},\bar{\Omega}\rbrace$, the first term in \eqref{eq:hodge-norm} is the contribution from its $(3,0)\oplus(0,3)$ part and the second that of its $(2,1)\oplus(1,2)$ part. We refer to appendix \ref{app:hodge-theory} for the derivation of this formula.

\noindent As for any quadratic form, there is a symmetric matrix $\mathcal{M}$ defining the norm $\norm{v}_{\mathcal{M}}^2=v^{\alpha}\mathcal{M}_{\alpha\beta}v^{\beta}$, where $v^{\alpha}=(v_{I},v^{J})$ so that $\alpha=1,\ldots,2h^{2,1}+2$. The components of $\mathcal{M}$ can be extracted from the RHS of \eqref{eq:hodge-norm} and read
\begin{equation}\label{eq:hodge-matrix}
\mathcal{M}_{\alpha\beta}=2e^{K_{\text{cs}}}\text{Re}\left((\eta\mathbf{\Pi})_{\alpha}\overline{(\eta\mathbf{\Pi})}_{\beta}+G^{i\bar{j}}(\eta D_{i}\mathbf{\Pi})_{\alpha}\overline{(\eta D_{j}\mathbf{\Pi})}_{\beta}\right).
\end{equation}
Note that it is sufficient to consider the real part of the expression inside the brackets, as any imaginary part will cancel out when contracting the matrix with real vectors $v^{\alpha}$. The corresponding inverse diagonal components are related to the original ones through symplecticity of $\mathcal{M}$, giving $\mathcal{M}^{\alpha+h^{2,1}+1,\hspace{0.1mm}\alpha+h^{2,1}+1}=\mathcal{M}_{\alpha\alpha}$ for any $\alpha=1,\ldots,h^{2,1}+1$. Consider now a vacuum solution in the sense that it satisfies $D_{\tau}W=0$ where
\begin{equation}
    D_{\tau}W=\partial_{\tau}W+(\partial_{\tau}K)W=-\int_{Y}\left(H_{3}+\frac{F_{3}-\tau H_{3}}{\tau-\bar{\tau}}\right)\wedge\Omega=-\frac{1}{\tau-\bar{\tau}}\int_{Y}(F_{3}-\bar{\tau}H_{3})\wedge\Omega=0,
\end{equation}
which can be written in components as $(\mathbf{f}-\bar{\tau}\mathbf{h})^{T}\eta\mathbf{\Pi}=0$. Using this vacuum constraint, the flux superpotential \eqref{eq:general_W0} can be expressed in terms of just the NSNS flux,
\begin{equation}\label{eq:vacuum-superpotential-h-flux}
    W=(\mathbf{f}-\tau\mathbf{h})^{T}\eta\mathbf{\Pi}=(\bar{\tau}-\tau)\mathbf{h}^{T}\eta\mathbf{\Pi}=-2i\hspace{0.3mm}\tau_{2}\hspace{0.3mm}\mathbf{h}^{T}\eta\mathbf{\Pi}.
\end{equation}
Consequently, the gauge invariant superpotential can be expressed as
\begin{equation}\label{eq:value-superpotential-norms}
    \abs{W_{0}}^2=\frac{4}{\pi}\tau_{2}\hspace{0.3mm}e^{K_{\text{cs}}}\abs{\mathbf{h}^{T}\eta\mathbf{\Pi}}^2,
\end{equation}
after substituting the Kähler potential \eqref{eq:Kahler-potential}. At the vacuum, the flux charge \eqref{eq:tadpole} can be written just in terms of the NSNS fluxes, namely
\begin{equation}\label{eq:identity-Nflux-hodgenorm}
    N_{\text{flux}}=\mathbf{f}^{T}\eta\mathbf{h}=\mathbf{h}^{T}(\tau_{1}\mathbb{I}+\tau_{2}\mathcal{M}^{T}\eta^{T})\eta\mathbf{h}=\tau_{2}\mathbf{h}^{T}\mathcal{M}\mathbf{h}=\tau_{2}\norm{H_{3}}_{\mathcal{M}}^2,
\end{equation}
where in the middle step we have inserted the vacuum equation \eqref{eq:vacua-condition}. Now, substituting the value of $\tau_{2}$ from \eqref{eq:identity-Nflux-hodgenorm} in \eqref{eq:value-superpotential-norms} and using the Hodge norm expression in \eqref{eq:hodge-norm} for the $H_{3}$ flux, one arrives at
\begin{equation}\label{eq:exact-superpotential-identity}
    \abs{W_{0}}^2=\frac{2}{\pi}N_{\text{flux}}f_{H}(z),\qquad f_{H}(z)=\frac{\abs{\mathbf{h}^{T}\eta\mathbf{\Pi}}^2}{\abs{\mathbf{h}^{T}\eta\mathbf{\Pi}}^2+G^{i\bar{j}}(\mathbf{h}^{T}\eta D_{i}\mathbf{\Pi})\overline{(\mathbf{h}^{T}\eta D_{j}\mathbf{\Pi})}}\in[0,1].
\end{equation}
The factor $f_{H}$ here is the $(3,0)\oplus(0,3)$ fraction of the Hodge norm of the NSNS flux. Since $f_{H}\leq 1$, \eqref{eq:exact-superpotential-identity} reproduces the bound $\abs{W_{0}}^2\leq\frac{2}{\pi}N_{\text{flux}}$ from \cite{Denef:2004ze}, and it shows that a small superpotential requires an NSNS flux which is almost purely of type $(2,1)\oplus(1,2)$ at the vacuum.

Let us turn to the connection between the actual value of the vacuum superpotential $\abs{W_{0}}$ and the associated Fisher eigenvalue. Under a rescaling of the fluxes, the flux superpotential $W$ is a homogeneous function of degree $1$ due to its linear dependence on $\theta$ in \eqref{eq:W0_linear}. Since the Kähler potential does not depend explicitly on $\theta$, one has that $\abs{W_{0}}$ is also homogeneous of degree $1$. This homogeneity allows us to write
\begin{equation}
\abs{W_{0}}=\theta^{A}\partial_{A}\abs{W_{0}}=\boldsymbol{\theta}^{T}\nabla_{\theta}\abs{W_{0}}.
\end{equation}
By means of the Cauchy-Schwarz inequality, it then follows that
\begin{equation}
\abs{W_{0}}\leq\norm{\boldsymbol{\theta}}\norm{\nabla_{\theta}\abs{W_{0}}}=\norm{\boldsymbol{\theta}}\sqrt{\lambda},
\end{equation}
with $\norm{\cdot}$ being the Euclidean norm. Put differently, the Fisher eigenvalue controls the maximal size of the superpotential: for fixed $\|\boldsymbol{\theta}\|$, small $\lambda$ forces $|W_0|$ to be
small, while large $\lambda$ does not preclude small $|W_0|$.

Before analysing the asymptotic behaviour of the Fisher eigenvalue, we make a digression to show how flux quantisation, combined with the structure of the Hodge matrix $\mathcal{M}_{\alpha\beta}$, restricts the active flux components in certain regions of moduli space.

\subsection{Integrality of fluxes and depth in moduli space}\label{sec:fine-tuning-and-distance}

So far, we have not really used the fact that both NSNS and RR fluxes are integrally quantised, lying in the integral cohomology group $H^{3}(Y,\mathbb{Z})$. We show that this quantisation requirement is quite stringent, and forbids switching on some flux components in certain regions of the moduli space in order to be consistent with the tadpole bound. 

To start off, recall that the Kähler potential in $\mathcal{M}_{\text{cs}}$ is computed from the prepotential $\mathcal{F}$ in \eqref{eq:general_prepotential} through \cite{Grimm:2005fa}
\begin{equation}\label{Kahler-metric-from-prepotential}
    K_{\text{cs}}=-\log\left[2i(\mathcal{F}-\bar{\mathcal{F}})-i(\partial_{j}\mathcal{F}+\partial_{\bar{j}}\bar{\mathcal{F}})(z^{j}-\bar{z}^{j})\right].
\end{equation}
If we now focus on the case of a prepotential containing a classical term\footnote{More precisely, the classical prepotential usually refers to the cubic term in $z$ in \eqref{eq:general_prepotential}. Nonetheless, it is not hard to see that since the coefficients $a_{ij}$ and $b_{i}$ are real, both the quadratic and linear terms in $z$ drop out in the RHS of \eqref{Kahler-metric-from-prepotential}.} and a constant term arising from perturbative $\alpha^{\prime}$ corrections in the mirror, i.e. we ignore the instanton corrections in \eqref{eq:general_prepotential}, the Kähler potential takes the known form
\begin{equation}\label{eq:Kahler-potential-cs-in-terms-of-mirror}
    K_{\text{cs}}=-\log\left(\frac{4}{3}\kappa_{ijk}y^{i}y^{j}y^{k}+\frac{\chi\hspace{0.5mm}\zeta(3)}{4\pi^3}\right),\qquad i=1,\ldots,h^{2,1}(Y).
\end{equation}
The first quantity inside the log is nothing but the usual term $8\mathcal{V}_{\tilde{Y}}$, with $\mathcal{V}_{\tilde{Y}}$ the volume of the mirror $\tilde{Y}$ and $y^{i}$ being the real mirror Kähler moduli. Also, recall that this Kähler potential (without the constant piece) admits an expression in terms of the period vector $\mathbf{\Pi}$ given by the last term in equation \eqref{eq:Kahler-potential}. 

Note that this object only depends on the imaginary parts of the moduli $z^{i}=x^{i}+iy^{i}$, as it should, since the axionic real parts $x^{i}$ are protected by a perturbative continuous shift symmetry, which gets broken to a discrete subgroup only in the presence of instanton corrections. Combined with the corresponding transformation of the fluxes, the monodromy
$x^i\to x^i+1$ allows one to restrict the axions to the interval $(-\frac12,\frac12]$, as in the canonical frame. For simplicity, let us just call the classical volume-like term $V_{c}$, with the subscript $c$ standing for classical, and the constant term $C$, that is
\begin{equation}
    V_{c}=\frac{4}{3}\kappa_{ijk}y^{i}y^{j}y^{k}, \qquad C=\frac{\chi\hspace{0.5mm}\zeta(3)}{4\pi^3}\approx 5.23,
\end{equation}
where we have evaluated the constant for the geometry at hand. The quantum corrected volume $V$ is the factor inside the log of \eqref{eq:Kahler-potential-cs-in-terms-of-mirror}, i.e. $V=V_{c}+C$. Building on the Kähler potential \eqref{eq:Kahler-potential-cs-in-terms-of-mirror}, the associated Kähler metric reads 
\begin{equation}\label{eq:Kahler-metric-from-kahler-sector}
    G_{ij}=\frac{1}{4}\left(\frac{V_{i}V_{j}}{V^2}-\frac{V_{ij}}{V}\right),\qquad V_{i}=\partial_{y^{i}}V.
\end{equation}
Since $V_{c}$ is homogeneous of degree $3$ in the variables $y^{i}$, it follows that
\begin{equation}\label{eq:homogeneity-eqs}
    y^{i}V_{i}=3V_{c},\qquad V_{ij}y^{j}=2V_{i}.
\end{equation}
We also record two further identities which again follow from homogeneity of $V_{c}$ and whose exact derivation can be found in appendix \ref{app:homogeneity-identities}. These are, 
\begin{equation}\label{eq:homogeneity-identities}
    G_{ij}y^{j}=\frac{V_{c}-2C}{4V^2}V_{i},\qquad V_{i}G^{ij}V_{j}=\frac{12V_{c}V^2}{V_{c}-2C},
\end{equation}
where $G^{ij}$ is the inverse metric of $G_{ij}$. The latter two equations are useful when computing components of the Hodge matrix. Let us consider the first diagonal component of \eqref{eq:hodge-matrix}
\begin{equation}\label{eq:1,1-eq-1}
    \mathcal{M}_{11}=\frac{2}{V}\left(1+G^{i\bar{j}}\partial_{i}K_{\text{cs}}\overline{\partial_{j}K_{\text{cs}}}\right).
\end{equation}
This follows from substituting $(\eta\mathbf{\Pi})_{1}=X^{0}=1$ together with $(\eta D_{i}\mathbf{\Pi})_{1}=\partial_{i}K_{\text{cs}}$. Since $K_{\text{cs}}$ just depends on $y^{i}$, we can phrase it in terms of the real derivatives $\partial_{i}=\partial_{y^{i}}$ as follows
\begin{equation}\label{eq:derivation-M11}
    \mathcal{M}_{11}=\frac{2}{V}\left(1+\frac{G^{ij}V_{i}V_{j}}{4V^2}\right)=\frac{2}{V}\frac{4V_{c}-2C}{V_{c}-2C}=\frac{4(2V-3C)}{V(V-3C)}, \qquad \partial_{i}K_{\text{cs}}=\frac{iV_{i}}{2V},
\end{equation}
where we have used the second relation in \eqref{eq:homogeneity-identities}. The Cauchy-Schwarz inequality in the Hodge metric gives
\begin{equation}\label{eq:hodge-bound-H3}
    \norm{H_{3}}_{\mathcal{M}}^2=\mathbf{h}^{T}\mathcal{M}\mathbf{h}\geq\frac{(h^{\alpha})^2}{\mathcal{M}^{\alpha\alpha}}.
\end{equation}
Combining it with \eqref{eq:identity-Nflux-hodgenorm}, we find a lower bound for the flux charge in terms of the $\alpha$-th component of both the NSNS flux vector and the Hodge matrix given by
\begin{equation}\label{eq:integrality-condition-1}
    N_{\text{flux}}\geq\frac{\tau_{2}}{\mathcal{M}^{\alpha\alpha}}(h^{\alpha})^2,
\end{equation}
which holds for any $\alpha$. Let us begin by considering the case $\alpha=4$. Integrality of the fluxes implies that if $h^{4}\neq 0$, then
\begin{equation}\label{eq:bound-integrality-1}
    N_{\text{flux}}\geq\tau_{2}\frac{V(V-3C)}{4(2V-3C)},
\end{equation}
obtained by invoking $\mathcal{M}^{44}=\mathcal{M}_{11}$ together with the explicit expression \eqref{eq:derivation-M11}. 

To explore the consequences of the lower bound in \eqref{eq:bound-integrality-1}, we restrict ourselves to a family of subregions in $\mathcal{M}_{\tau}\times\mathcal{M}_{\text{cs}}$ parametrised by $s$, with $s>1$, since otherwise the instantonic corrections that we have been ignoring so far are no longer negligible. We define them as
\begin{equation}
    U(s)=\lbrace (\tau,z^{1},z^{2})\in\mathcal{M}_{\tau}\times\mathcal{M}_{\text{cs}}\hspace{0.8mm}\rvert\hspace{0.8mm} y^{1},y^{2}\geq s,\hspace{1mm}\tau_{2}\geq 3\hspace{0.5mm}\text{max}(y^{1},y^{2})\rbrace.
\end{equation}
At this point, let us emphasise that there is nothing special about the factor of $3$ in the lower bound for $\tau_{2}$, and any other value could have been chosen. In our case, the main motivation behind this choice is our interest in weakly coupled vacua. Moreover, choosing a smaller value would require larger values of $s$ to violate the bound \eqref{eq:bound-integrality-1}, as we will see next. In fact, there is a trade-off between the factor appearing in the lower bound for $\tau_{2}$ and the depth variable in moduli space, which can be adjusted to achieve a violation of the bound \eqref{eq:bound-integrality-1}.

Observe that the RHS of \eqref{eq:bound-integrality-1} is increasing in $V$ when $V\geq 3C$ and in $\tau_{2}$. We point out that even when considering the lowest plausible value $s=1$, $V>3C$ holds for the geometry in \ref{sec:specialisation-geoemtry}, meaning that it attains its minimum for the smallest possible values of $V$ and $\tau_{2}$. Since all the intersection numbers $\kappa_{ijk}$ are non-negative for this geometry, $V$ is minimised when $y^{1}=y^{2}=s$. Taking then $\tau_{2}=3s$, one has that
\begin{equation}\label{eq:def-B4ss}
    N_{\text{flux}}\geq B_{\text{4}}(s,s)=\frac{3s(28s^3+C)(28s^{3}-2C)}{4(56s^{3}-C)},\qquad B_{4}(y^{1},y^{2})=\frac{3\hspace{0.5mm}\text{max}(y^{1},y^{2})}{\mathcal{M}_{11}}.
    \end{equation}
Moreover, recall from the tadpole bound discussion around equation \eqref{eq:tadpole-bound} that $N_{\text{flux}}\leq Q_{D3}$, which for the model under consideration takes the value $Q_{D3}=138$. However, already for $s=2$, we have that $B_{4}(2,2)=165.83$, contradicting therefore the tadpole bound. It is precisely the value $s=1.911$ that determines the threshold signalling the incompatibility with the tadpole bound, as $B_{4}(1.911,1.911)\simeq138$. Hence, vacua in the regions $U(s\gtrsim1.911)$ must have a vanishing fourth component of the NSNS flux, that is $h^{4}=h_{2}^{0}=0$. 

A similar analysis applies to the other flux components. For $\alpha=5,6$, the bound \eqref{eq:hodge-bound-H3} requires knowledge of $\mathcal{M}^{55}=\mathcal{M}_{22}$ and $\mathcal{M}^{66}=\mathcal{M}_{33}$. Let us then derive the components $\mathcal{M}_{k+1,l+1}$ with $k,l=1,2$, from the general expression \eqref{eq:hodge-matrix}. By definition,
\begin{equation}\label{eq:matrix-k-l}
    \mathcal{M}_{k+1,l+1}=2e^{K_{\text{cs}}}\text{Re}\left[z^{k}\bar{z}^{l}+G^{i\bar{j}}(\delta_{i}^{k}+(\partial_{i}K_{\text{cs}})z^{k})(\delta_{j}^{l}+(\overline{\partial_{j}K_{\text{cs}}})\bar{z}^{l})\right],
\end{equation}
with $(\eta\mathbf{\Pi})_{k+1}=X^{k}=z^{k}$ and $(\eta D_{i}\mathbf{\Pi})_{k+1}=\delta^{k}_{i}+(\partial_{i}K_{\text{cs}})z^{k}$. The cross terms inside the brackets can be further simplified by means of the Kähler derivatives in the second expression of \eqref{eq:derivation-M11}, yielding
\begin{equation}
    \text{Re}\left[G^{k\bar{j}}(\overline{\partial_{j}K_{\text{cs}}})\bar{z}^{l}+G^{i\bar{l}}(\partial_{i}K_{\text{cs}})z^{k}\right]=-G^{kj}\frac{V_{j}}{2V}y^{l}-G^{il}\frac{V_{i}}{2V}y^{k}.
\end{equation}
The contractions of the form $G^{kj}V_{j}$ can be obtained by acting on the first identity in \eqref{eq:homogeneity-identities} with the inverse metric, giving rise to
\begin{equation}\label{eq:contraction-with-inverse}
    G^{kj}V_{j}=\frac{4V^2}{V_{c}-2C}y^{k}.
\end{equation}
Inserting this into the matrix \eqref{eq:matrix-k-l} and using the second identity in \eqref{eq:homogeneity-identities} for the term proportional to $z^{k}\bar{z}^{l}$ inside the brackets, one arrives at
\begin{equation}\label{eq:M-k+1-l+1}
    \mathcal{M}_{k+1,l+1}=\frac{2}{V}\left[G^{kl}+\frac{4V_{c}-2C}{V_{c}-2C}x^{k}x^{l}-\frac{6C}{V_{c}-2C}y^{k}y^{l}\right].
\end{equation}
Note that in contrast with $\mathcal{M}_{11}$, these components depend both on saxions $y^{i}$ and on axions $x^{i}$. In particular, the diagonal components take the expressions
\begin{equation}\label{eq:diagonal-components-hodge-matrix}
    \mathcal{M}_{22}=\frac{4(y^1)^2}{V}+\frac{y^1}{y^2(3y^1+y^2)}+\frac{2}{V}\frac{4V_{c}-2C}{V_{c}-2C}(x^{1})^2,\quad \mathcal{M}_{33}=\frac{4(y^2)^2}{V}+\frac{3}{y^2}+\frac{2}{V}\frac{4V_{c}-2C}{V_{c}-2C}(x^{2})^2,
\end{equation}
whose derivation can be found in appendix \ref{app:hodge-theory}. Now, similarly to the $\alpha=4$ discussion, in the event of non-vanishing components $h^{5}=h_{2}^{1}$ and $h^{6}=h_{2}^{2}$, the RHS of \eqref{eq:integrality-condition-1} is minimised for $\tau_{2}=3\hspace{0.5mm}\text{max}(y^{1},y^{2})$ and $\lvert x^{i}\rvert=1/2$. Following the same steps as for the previous case, we define
\begin{equation}\label{eq:B5-B6-functions}
    B_{5}(y^{1},y^{2})=\frac{3\hspace{0.5mm}\text{max}(y^{1},y^{2})}{\mathcal{M}_{22}},\qquad B_{6}(y^{1},y^{2})=\frac{3\hspace{0.5mm}\text{max}(y^{1},y^{2})}{\mathcal{M}_{33}}.
\end{equation}
One can show that these functions attain their minima along the diagonal $y^{1}=y^{2}=s$, that is $B_{5,6}(y^{1},y^{2})\geq B_{5,6}(s,s)$, see appendix \ref{app:hodge-theory}. Then, the bound in \eqref{eq:integrality-condition-1} is nothing but $Q_{D3}\geq B_{5,6}(s,s)$, with
\begin{equation}
    B_{5}(s,s)=\frac{3s}{\frac{4s^2}{28s^{3}+C}+\frac{1}{4s}+\frac{56s^{3}-C}{(28s^{3}+C)(28s^{3}-2C)}},\quad B_{6}(s,s)=\frac{3s}{\frac{4s^2}{28s^{3}+C}+\frac{3}{s}+\frac{56s^{3}-C}{(28s^{3}+C)(28s^{3}-2C)}}.
\end{equation}
It turns out that the inequality for $B_{5}$ is saturated when $s\simeq 4.27$, which means that  $h^{5}=h_{2}^{1}=0$ across the region $U(s\gtrsim 4.27)$. In the case of $B_{6}$, this happens when $s\simeq 12.025$, fixing $h^{6}=h_{2}^{2}=0$ across $U(s\gtrsim 12.025)$.  

So far, our analysis has focused exclusively on the NSNS fluxes. However, an analogous argument applies to the RR sector, as they are related to the former through the vacuum equations. More concretely, at any vacuum solution, the Hodge norm for the RR flux can be written in terms of the NSNS one by means of the ISD condition \eqref{eq:vacua-condition}, giving 
\begin{equation}
    \norm{F_{3}}_{\mathcal{M}}^2=\mathbf{f}^{T}\mathcal{M}\mathbf{f}=\tau_{1}^2\mathbf{h}^{T}\mathcal{M}\mathbf{h}+2\tau_{1}\tau_{2}\mathbf{h}^{T}\mathcal{M}\eta^{T}\mathcal{M}\mathbf{h}+\tau_{2}^2\mathbf{h}^{T}\mathcal{M}\eta^{T}\mathcal{M}\eta\mathcal{M}\mathbf{h}.
\end{equation}
Due to symplecticity, one has that $\mathcal{M}\eta^{T}\mathcal{M}=\eta^{T}$, which implies the vanishing of the second term by symmetry arguments. By the same token, one has that $\mathcal{M}\eta^{T}\mathcal{M}\eta=\mathbb{I}$, so that the first and third terms can be merged into one, yielding
\begin{equation}
    \norm{F_{3}}_{\mathcal{M}}^2=\abs{\tau}^2\mathbf{h}^{T}\mathcal{M}\mathbf{h}=\frac{\abs{\tau}^2N_{\text{flux}}}{\tau_{2}},
\end{equation}
where in the last step we invoked \eqref{eq:identity-Nflux-hodgenorm}. Finally, analogously to \eqref{eq:hodge-bound-H3}, we apply the Cauchy-Schwarz inequality to obtain the bound
\begin{equation}\label{eq:hodge-bound-F3}
    N_{\text{flux}}\geq \frac{\tau_{2}}{\abs{\tau}^2\mathcal{M}^{\alpha\alpha}}(f^{\alpha})^2,
\end{equation}
which holds for any $\alpha$. We can now proceed as for the NSNS fluxes. In the case of $\alpha=4$, and again using that $\mathcal{M}^{44}=\mathcal{M}_{11}$, one has for $f^{4}=f^{0}_{2}\neq 0$ the following bound 
\begin{equation}\label{eq:upper-bound-tau2}
    N_{\text{flux}}\geq \frac{\tau_{2}}{\frac{1}{4}+\tau_{2}^2}\frac{1}{\mathcal{M}_{11}}.
\end{equation}
In contrast to the NSNS flux bound \eqref{eq:bound-integrality-1}, the RHS here decreases as we go to weaker couplings. Thus, the exclusion of $f^{0}_{2}\neq 0$ would require an upper bound on $\tau_{2}$. To see how this works, let us define $b(y^1,y^2)=Q_{D3}\,\mathcal{M}_{11}(y^1,y^2)$ and state the following claim: 
\begin{equation}\label{eq:claim} 
b<1\hspace{2mm}\text{and}\hspace{2mm} \tfrac12<\tau_{2}<\tau_{2}^{*}(y^1,y^2)=\frac{1+\sqrt{1-b^2}}{2b} \;\Rightarrow\; f^{0}_{2}=0. 
\end{equation} 
One readily proves this by contradiction. Assuming $f_{2}^{0}\neq 0$, the bound \eqref{eq:upper-bound-tau2} together with $N_{\text{flux}}\leq Q_{D3}$ gives 
\begin{equation} 
b=Q_{D3}\mathcal{M}_{11}\geq N_{\text{flux}}\mathcal{M}_{11} \geq\frac{\tau_{2}}{\frac{1}{4}+\tau_{2}^2} >\frac{\tau^{*}_{2}}{\frac{1}{4}+(\tau^{*}_{2})^2}=b, 
\end{equation} where the strict inequality uses that $\tau_2/(\frac14+\tau_2^2)$ is decreasing for $\tau_2>\frac12$, which holds throughout $U(s)$. The condition $b<1$ enforces $\mathcal{M}_{11}<1/Q_{D3}$ and in turn, since $\mathcal{M}_{11}$ is maximised when $y^{1}=y^{2}=s$ due to the non-negativity of $\kappa_{ijk}$, this translates into
\begin{equation}
    \frac{3s}{B_{4}(s,s)}<\frac{1}{Q_{D3}},
\end{equation}
with $B_{4}(s,s)$ defined in \eqref{eq:def-B4ss}. By solving the inequality we find that $s\gtrsim 3.406$, which means that $f_{2}^{0}$ must vanish across the patch $\tilde{U}(s\gtrsim 3.406)$, with
\begin{equation}\label{eq:special-branch}
    \tilde{U}(s)=\lbrace (\tau,z^{1},z^{2})\in\mathcal{M}_{\tau}\times\mathcal{M}_{\text{cs}}\hspace{0.8mm}\rvert\hspace{0.8mm} y^{1},y^{2}\geq s,\hspace{1mm}\tau_{2}^{*}\geq\tau_{2}\geq 3\hspace{0.5mm}\text{max}(y^{1},y^{2})\rbrace.
\end{equation}
Note that $\tilde{U}(s)$ is nothing but $U(s)$ with an upper bound on $\tau_{2}$ to exclude the weakest couplings. Combining the results for both NSNS and RR fluxes, we conclude that across $\tilde{U}(s\gtrsim 12.025)$, there is a special flux branch for which some components vanish, $f^{0}_{2}=h_{2}^{0}=h^{1}_{2}=h^{2}_{2}=0$. Equivalently, the flux lattice truncates to a lower-dimensional one across the region $\tilde{U}(s\gtrsim 12.025)$, where $\mathbb{Z}^{12}\rightarrow\mathbb{Z}^{8}$.

Although our analysis has been carried out in a specific model, the mechanism behind this truncation is general. Near the boundaries of complex structure moduli space, the Hodge norm of fluxes with large enough weight under the limiting mixed Hodge structure grows with the saxions \cite{Schmid:1973cdo,Cattani:1986,Grimm:2018ohb,Grimm:2018cpv}. Combined with flux integrality, this makes such components increasingly dominant in the D3-brane tadpole. This effect underlies the systematic study of flux vacua in asymptotic regions \cite{Grimm:2019ixq,Grimm:2021ckh} and the tadpole bounds at large complex structure and in asymptotic limits \cite{Plauschinn:2021hkp,Grana:2022dfw}, which support the tadpole conjecture \cite{Bena:2020xrh,Bena:2021wyr}. The details, however, are model dependent: which components are truncated, and the thresholds at which this happens, depend on the intersection numbers, on the tadpole bound, on the region of moduli space considered and, in particular, on how the string coupling scales relative to the complex structure saxions. For a general analysis of flux vacua at large complex structure and weak coupling, including the interplay between the tadpole bound and the range of the saxion vevs, see \cite{Marchesano:2021gyv}.

Along these lines and using the same model, a recent work \cite{Chauhan:2026gid} used Topological Data Analysis (TDA) techniques to study which flux directions exhibit the highest concentration of variance, indicating an effective reduction of the flux space to just those dominant components. Our discussion here can be seen as an analytic version of the flux space reduction found in their empirical analysis.

\subsection{Fine-tuning at large complex structure and weak coupling}\label{sec:fine-tuning-and-string-coupling}
Our objective now is to study the asymptotic dependence of $\lambda$ on both $z^{i}$ and $\tau$ when moving to regions where $y^{i}\gg 1$ and $\tau_{2}\gg 1$, which corresponds to the deep LCS and weak-coupling limit, $g_{s}\ll 1$. To this end we study the asymptotic behaviour of the functions $G(z)$, $D(\tau)$ and $\mathcal{A}$ defined in \eqref{eq:functions-composing-lambda}. Clearly the function $D(\tau)$ approaches the limit
\begin{equation}\label{eq:D-limit}
    D(\tau)=\frac{1+\abs{\tau}^2}{\tau_{2}}\xrightarrow{\tau_{2}\gg 1}\tau_{2}.
\end{equation}
As for the bounded function $\mathcal{A}$, its $\tau$ dependence drops out in the weak coupling limit, as can be seen by looking at its $\varepsilon$-like piece
\begin{equation}\label{eq:A-limit}
    \frac{\mathbf{\Xi}^{T}\mathbf{\Xi}}{\norm{\mathbf{\Xi}}^2}=\frac{(1+\tau^2)\mathbf{\Pi}^{T}\mathbf{\Pi}}{(1+\abs{\tau}^2)\norm{\mathbf{\Pi}}^2}\xrightarrow{\tau_{2}\gg 1}-\frac{\tau_{2}^2\hspace{0.5mm}\mathbf{\Pi}^{T}\mathbf{\Pi}}{\tau_{2}^2\hspace{0.5mm}\norm{\mathbf{\Pi}}^2}=-\frac{\mathbf{\Pi}^{T}\mathbf{\Pi}}{\norm{\mathbf{\Pi}}^2}.
\end{equation}
Hence, at fixed complex structure moduli and over the full flux lattice, $\lambda$ grows linearly with $\tau_2$, i.e.\ inversely with the string coupling. Moreover, when considering both limits $\tau_{2}\gg 1$ and $y^{i}\gg 1$ simultaneously, we find that 
\begin{equation}
    \frac{\mathbf{\Xi}^{T}\mathbf{\Xi}}{\norm{\mathbf{\Xi}}^2}\xrightarrow{\tau_{2}\gg 1}-\frac{\mathbf{\Pi}^{T}\mathbf{\Pi}}{\norm{\mathbf{\Pi}}^2} \xrightarrow{y^{i}\gg 1} 1,
\end{equation}
which tells us that the bounded function $\mathcal{A}$ only depends on the phase of the flux superpotential $W$ via\footnote{As an aside, at the strong coupling enhanced symmetry point $\tau=i$, the LHS of \eqref{eq:A-limit} vanishes automatically and thus $\mathcal{A}$ drops its dependence on the phase and stabilises to $\mathcal{A}\rightarrow 1$ regardless of the complex structure moduli space location.}
\begin{equation}\label{eq:A-component}
    \mathcal{A}\simeq 1+\text{Re}(u^2)=1+\cos 2\varphi.
\end{equation}

Regarding the asymptotic behaviour of $G(z)$, we begin by noting that in this limit the period vector norm scales like
\begin{equation}
    \norm{\mathbf{\Pi}}^2\simeq\abs{\mathcal{F}_{0}}^2\simeq\frac{1}{64}e^{-2K_{\text{cs}}},
\end{equation}
since the period vector in \eqref{eq:period-vector} is dominated by its leading component $\mathcal{F}_{0}$, as it scales cubically with the depth variables
\begin{equation}
    \mathcal{F}_{0}\xrightarrow{y^{i}\gg 1}\frac{-i}{6}\kappa_{ijk}y^{i}y^{j}y^{k},
\end{equation}
together with the fact that the Kähler potential \eqref{eq:Kahler-potential-cs-in-terms-of-mirror} can be just considered classical in this limit, neglecting the constant term inside the logarithm. The function $G(z)$ then goes like
\begin{equation}
    G(z)\xrightarrow{y^{i}\gg 1} \frac{1}{256}e^{-K_{\text{cs}}}.
\end{equation}
Putting everything together, in the deep LCS region and weak coupling limit, the fine-tuning eigenvalue takes the following asymptotic behaviour,
\begin{equation}\label{eq:lambda-asymptotic}
    \lambda\simeq \frac{1}{128\pi}(1+\cos 2\varphi)\hspace{0.5mm}\tau_{2}\hspace{0.5mm}e^{-K_{\text{cs}}}.
\end{equation}
In particular, all $z^{i}$ dependence is encapsulated in the factor $e^{-K_{\text{cs}}}\simeq V_{c}$. Since $V_c$ is proportional to the volume of the mirror, it is positive inside the K\"ahler cone and homogeneous of degree three, so it grows cubically along any ray deep into the LCS region. In our example all intersection numbers are non-negative and $\partial V_c/\partial y^i=4\kappa_{ijk}y^jy^k>0$ for each $i$, so $V_c$ moreover increases monotonically in each $y^i$ separately. In the canonical frame, the Fisher eigenvalue would thus grow in the asymptotic weak-coupling and deep LCS regime, rendering such vacua less natural. 

However, one ingredient is still missing at this point, namely the truncation of the flux lattice imposed by the tadpole bound. In essence, as one moves to weaker coupling and deeper into the LCS region, the inequalities \eqref{eq:integrality-condition-1} and \eqref{eq:hodge-bound-F3} force certain flux components to vanish, thereby truncating the parameter space to a lower-dimensional one,
$\mathcal{P}\rightarrow\mathcal{P}_{t}$. Variations along the excluded directions do not connect to admissible vacua, so that the Fisher matrix should be restricted to $\mathcal{P}_{t}$. While the restricted Fisher matrix still has a single non-trivial eigenvalue, the factorisation of $\lambda$ in \eqref{eq:lambda-decomposition} generally breaks down, since the admissible RR and NSNS directions need not coincide. Recalling \eqref{eq:lambda-mid-decomposition}, the truncated eigenvalue reads
\begin{equation}\label{eq:truncated-lambda}
    \lambda_{t}=\frac{e^{K_{\text{cs}}}}{2\pi\tau_{2}}\,\norm{\boldsymbol{\Xi}_{t}}^2\mathcal{A}_{t}\,,
    \qquad \mathcal{A}_{t}=1+\frac{\operatorname{Re}\big(u^2\,\boldsymbol{\Xi}_{t}^{T}\boldsymbol{\Xi}_{t}\big)}{\norm{\boldsymbol{\Xi}_{t}}^2}\,,
\end{equation}
where $\boldsymbol{\Xi}_{t}$ consists of the components of $\boldsymbol{\Xi}$ that multiply, through \eqref{eq:W0_linear}, the flux components not excluded by the tadpole bound, and the subscript $t$ stands for truncated. The bounded factor $\mathcal{A}_{t}\in[0,2]$ is defined as before. In other words, the asymptotic behaviour \eqref{eq:lambda-asymptotic} holds when variations along all flux directions are considered, spanning the full twelve-dimensional
lattice, but not necessarily when some of these directions are excluded. In that case, the behaviour of $\lambda_{t}$ depends on the periods that the surviving flux directions couple to. To illustrate the impact of the tadpole bound, we next show for the special flux branch, how the growth of $\lambda$ given by \eqref{eq:lambda-asymptotic}, cubic in $y^{i}$ and linear
in $\tau_{2}$, is replaced by a constant or slower growth.

\subsubsection{Restriction to the special branch}

We conclude our analysis by computing the eigenvalue restricted to the admissible directions on the special flux branch of section \ref{sec:fine-tuning-and-distance}. On this branch $h_2^0=h_2^1=h_2^2=0$, so that the vacuum superpotential \eqref{eq:vacuum-superpotential-h-flux} reduces to
\begin{equation}
    W=-2i\tau_{2}(h_{1,0}+h_{1,1}z^{1}+h_{1,2}z^{2}).
\end{equation}
Taking the fourth component of equation \eqref{eq:vacua-condition} gives
\begin{equation}\label{eq:component-f4}
    f^{0}_{2}=-\tau_{2}\left(\mathcal{M}_{11}h_{1,0}+\mathcal{M}_{12}h_{1,1}+\mathcal{M}_{13}h_{1,2}\right).
\end{equation}
The off-diagonal components $\mathcal{M}_{1,l+1}$ are computed directly from the Hodge matrix \eqref{eq:hodge-matrix}. These are
\begin{equation}
    \mathcal{M}_{1,l+1}=2e^{K_{\text{cs}}}\text{Re}\left[\bar{z}^{l}+G^{i\bar{j}}(\partial_{i}K_{\text{cs}})(\delta_{j}^{l}+(\overline{\partial_{j}K_{\text{cs}}})\bar{z}^{l})\right], \qquad l=1,2.
\end{equation}
From \eqref{eq:derivation-M11}, $\partial_{i}K_{\text{cs}}$ is purely imaginary so the second term inside the square brackets vanishes and since $\partial_{i}K_{\text{cs}}\overline{\partial_{j}K_{\text{cs}}}$ is real, it follows that
\begin{equation}
\mathcal{M}_{1,l+1}=2e^{K_{\text{cs}}}x^{l}\left[1+G^{i\bar{j}}\partial_{i}K_{\text{cs}}\overline{\partial_{j}K_{\text{cs}}}\right]=\mathcal{M}_{11}x^{l}.
\end{equation}
Inserting these components back in \eqref{eq:component-f4} we can extract $\mathcal{M}_{11}$ as a common factor, yielding
\begin{equation}
    f^{0}_{2}=-\tau_{2}\mathcal{M}_{11}\left(h_{1,0}+h_{1,1}x^{1}+h_{1,2}x^{2}\right).
\end{equation}
Note that the vanishing of $f^{0}_{2}$ is directly linked to the vanishing of Im$W$, since from the two equations above Im$W=2f^{0}_{2}/\mathcal{M}_{11}$. This means that along the special branch, the flux superpotential is real, i.e. $\varphi\in\lbrace0,\pi\rbrace$ and $u^{2}=1$. The admissible flux directions are the eight directions $(f_{1,I},f_2^i,h_{1,I})$, which couple to the truncated enlarged period vector
$\boldsymbol{\Xi}^{T}_{t}=(X^I,-\mathcal{F}_i,-\tau X^I)$. Deep in the LCS region at
weak coupling, as is the case across $\tilde{U}(s\gtrsim 12.025)$, the dominant periods in $\boldsymbol{\Xi}_{t}$ are $\mathcal{F}_{i}\simeq\frac{1}{2}\kappa_{ijk}y^{j}y^{k}$ and
$\tau X^{i}\simeq -\tau_{2}y^{i}$, which couple to the flux directions $f_{2}^{i}$ and $h_{1,i}$, respectively. Since both are real at leading order, one has $\norm{\boldsymbol{\Xi}_{t}}^2\simeq\boldsymbol{\Xi}_{t}^{T}\boldsymbol{\Xi}_{t}$.
Together with the reality of the superpotential on this branch, meaning $u^2=1$, this implies that $\mathcal{A}_{t}$ in \eqref{eq:truncated-lambda} approaches its maximum value, $\mathcal{A}_{t}\rightarrow 2$. The truncated eigenvalue then takes the form
\begin{equation}\label{eq:lambdat}
\lambda_t\simeq\frac{1}{\pi V_{c}}\left(\tau_2\,\delta_{ij}\,y^iy^j
  +\frac{1}{4\tau_2}\,\delta^{il}\kappa_{ijk}\kappa_{lmn}\,y^jy^ky^my^n\right)
  \xrightarrow{\;y^i=y\;}\;
  \frac{1}{28\pi}\left(\frac{2\tau_2}{y}+\frac{281}{4}\,\frac{y}{\tau_2}\right),
\end{equation}
where we have approximated $V\simeq V_{c}$ and, for clarity, evaluated $\lambda_{t}$ along the diagonal $y^{i}=y$. Unlike \eqref{eq:lambda-asymptotic}, which along the diagonal grows as $\lambda\sim\tau_{2}y^{3}$, the truncated eigenvalue $\lambda_t$ is invariant under a common rescaling of $\tau_2$ and $y$. More concretely, it does not grow along rays of fixed $\tau_2/y$, where it approaches a constant, while it grows at most linearly if $y$ or $\tau_2$ is increased separately. 

Let us finish by mentioning that the flux component $f_{2}^{0}$ is not only responsible for moving the phase of the superpotential, but also sets a lower bound on the possible values of $\abs{W_{0}}$. This can be seen by looking directly at \eqref{eq:value-superpotential-norms}, i.e.
\begin{equation}
    \abs{W_{0}}^2=\frac{4\tau_{2}}{\pi V}\left((h_{1,1}y^{1}+h_{1,2}y^2)^2+\frac{(f_{2}^{0})^2}{\tau_{2}^2\mathcal{M}_{11}^2}\right)\geq \frac{4}{\pi V}\frac{(f_{2}^{0})^2}{\tau_{2}\mathcal{M}_{11}^2}\simeq \frac{(f_{2}^{0})^2 V}{16\pi\tau_{2}},
\end{equation}
using the asymptotic behaviour $\mathcal{M}_{11}\simeq 8/V$ valid for $V\gg C$. This bound of course does not set any constraint when working on the special branch, where $f_{2}^{0}=0$, and precludes the existence of exponentially small $\abs{W_{0}}$ for any non-trivial $f_{2}^{0}\neq 0$.

\section{Conclusions}\label{sec:conclusions}
In this work, we have applied a measure of naturalness or fine-tuning rooted in Fisher information theory \cite{Halverson:2026cxe} to the study of the flux landscape arising from Type IIB compactifications. In our construction, we identify the parameter space with the space of NSNS-RR fluxes and the observable of interest with the gauge-invariant flux superpotential $\abs{W_{0}}$.

As an example to illustrate our construction, we have considered a $\mathbb{Z}_{2}$ orientifold of a degree-18 hypersurface in the weighted projective space $\mathbb{P}^{4}(1,1,1,6,9)$. This manifold contains a large number of complex structure moduli, which are reduced to only $2$ along a specific symmetric locus. Restricting to this symmetric submanifold significantly reduces the dimensionality of both the moduli space and the flux lattice, providing a tractable setting in which to explore the notion of fine-tuning.

The degree of fine-tuning is determined by the unique non-zero eigenvalue of the Fisher Information Metric. We have shown that this eigenvalue factorises into a piece depending on the axio-dilaton, a piece depending on the complex structure moduli, and a bounded factor depending on the phase of the superpotential, and we have studied its asymptotic behaviour towards weak
coupling and deep into the LCS region. In the canonical frame and over the full flux lattice, it grows linearly with the inverse string coupling and cubically with the complex structure saxions. However, flux integrality combined with the tadpole bound and certain relations from Hodge theory
forces some flux components to vanish across specific patches of moduli space, truncating the flux lattice to a lower-dimensional one. Variations along the excluded directions do not connect to admissible vacua, so that the Fisher eigenvalue must be restricted accordingly, $\lambda\rightarrow\lambda_t$, and its behaviour is then governed by the periods that couple to the surviving flux directions. The large periods that make $|W_0|$ most sensitive to a flux direction also give that direction the largest contribution to the flux tadpole, so that deep
in the LCS regime at weak coupling the most sensitive directions are precisely those that the tadpole bound renders inaccessible. We have made this explicit on a specific patch of the $\mathbb{Z}_2$ orientifold, the special flux branch, where the flux lattice truncates as
$\mathbb{Z}^{12}\rightarrow\mathbb{Z}^{8}$ and the restricted eigenvalue no longer grows with depth along rays of fixed $\tau_2/y^i$, approaching a constant value. Since the surviving directions, and hence the behaviour of $\lambda_t$, depend on the patch of moduli space and on how the string
coupling scales relative to the saxions, other regions require a careful study of their own.

Despite our focus on a particular example, let us emphasise that the type of analysis carried out in this work can be applied to any family of Calabi–Yau compactifications and their corresponding flux vacua. Extending the study of this fine-tuning measure to a broader range of examples may help us gain a better understanding of the distribution and geometric structure of vacua with small superpotential across moduli space. This could complement both classic statistical approaches \cite{Douglas:2003um} and recent data-driven methods for characterising the flux landscape, such as genetic algorithm searches \cite{Cole:2019enn}, the use of Topological Data Analysis techniques to identify certain topological features of the landscape \cite{Chauhan:2026gid}, the study of analytic and statistical structures in ensembles with a large number of moduli \cite{Greene:2013ida, Krippendorf:2022gcl}, and the investigation of how the inclusion of leading instanton corrections to the prepotential, which may compete with the classical polynomial contribution near the boundary of the LCS region, affects the multiplicity of flux vacua and gives rise to monodromy shifts \cite{Chauhan:2026dss}. We would also like to highlight a very recent work \cite{Hamada:2026cik}, in which the authors generated millions of new candidate ISD vacua for the same model considered in this paper by employing an optimisation algorithm to minimise the scalar potential.

In this first application we have considered a single observable, for which the Fisher matrix reduces to a rank one object. A natural extension is to include several observables at once, such as the real and imaginary parts of $W_0$, the string coupling or the moduli masses, for which the off-diagonal structure of the Fisher matrix encodes correlations between them. Lastly, it would be interesting to investigate the behaviour of the fine-tuning measure in the interior of moduli space, where a large number of flux vacua are expected to be found. This regime is considerably less understood than the LCS limit, as it requires control over non-perturbative corrections arising from instanton towers, as well as new effects associated with the vicinity of singularities. We leave these investigations for future work.
\newpage
\subsubsection*{Acknowledgements}
We would like to thank Jensen Paull for valuable comments and discussions. We have used AI tools to verify the internal consistency of our computations. This work was supported by the Science and Technology Facilities Council (STFC), with grant number UKRI1758. 

\appendix

\section{Homogeneity identities}\label{app:homogeneity-identities}
The first identity in \eqref{eq:homogeneity-identities} follows straightforwardly from the definition of the Kähler metric in \eqref{eq:Kahler-metric-from-kahler-sector} and the homogeneity conditions \eqref{eq:homogeneity-eqs}, since
\begin{equation}\label{eq:homogeneity-app-1}
    G_{ij}y^{j}=\frac{1}{4}\left(\frac{V_{i}V_{j}}{V^2}-\frac{V_{ij}}{V}\right)y^{j}=\frac{3V_{c}-2(V_{c}+C)}{4V^2}V_{i}=\frac{V_{c}-2C}{4V^2}V_{i}.
\end{equation}
Now, to derive the second identity in \eqref{eq:homogeneity-identities}, we contract both sides of \eqref{eq:homogeneity-app-1} with $G^{ki}V_{k}$. The LHS simply turns into 
\begin{equation}\label{eq:homogeneity-app-2}
    G^{ki}V_{k}G_{ij}y^{j}=y^{j}V_{j}=3V_{c},
\end{equation}
where we have used the first equation in \eqref{eq:homogeneity-eqs} in the last step. Regarding the RHS, one has
\begin{equation}\label{eq:homogeneity-app-3}
    \frac{V_{c}-2C}{4V^2}V_{i}G^{ik}V_{k}.
\end{equation}
Equating \eqref{eq:homogeneity-app-2} and \eqref{eq:homogeneity-app-3} we arrive at the desired identity
\begin{equation}
    V_{i}G^{ij}V_{j}=\frac{12V_{c}V^2}{V_{c}-2C},
\end{equation}
which is the second expression in \eqref{eq:homogeneity-identities}.

\section{Hodge theory review}\label{app:hodge-theory}
Let us now turn to the discussion of some Hodge theory basics. The middle cohomology group of a Calabi-Yau threefold $H^{3}(Y,\mathbb{C})$ admits a basis in terms of the $H^{p,q}$ subspaces, which is given by $\lbrace\Omega, D_{i}\Omega,\overline{D_{i}\Omega},\bar{\Omega}\rbrace$, where $\Omega$ and $D_{i}\Omega$ span $H^{3,0}$ and $H^{2,1}$ respectively. It is customary to define the symplectic product of 3-forms $u,v\in H^{3}(Y,\mathbb{C})$ as
\begin{equation}\label{eq:symplectic-product}
    \langle u,v\rangle=\int_{Y}u\wedge v=\mathbf{u}^{T}\eta\mathbf{v},\qquad\eta=\begin{pmatrix}
    0&& \mathbb{I}\\
    -\mathbb{I}&& 0
\end{pmatrix},
\end{equation}
where $\mathbf{u,v}$ are the coefficients of $u,v$ when expanded in the real symplectic basis $(\alpha_{I},\beta^{J})$  of $H^{3}(Y,\mathbb{R})$. Clearly, the only non-trivial pairings of basis elements come from the combinations
\begin{equation}\label{eq:products-forms}
    \langle\Omega,\bar{\Omega}\rangle=-ie^{-K_{\text{cs}}},\qquad \langle D_{i}\Omega,\overline{D_{j}\Omega}\rangle=iG_{i\bar{j}}e^{-K_{\text{cs}}},
\end{equation}
with $G_{i\bar{j}}$ being the Weil-Petersson metric. In addition to the symplectic product in \eqref{eq:symplectic-product}, we define the Hodge product of 3-forms as
\begin{equation}\label{eq:product-u-v}
    \langle u,v\rangle_{\mathcal{M}}=\int_{Y}u\wedge* \bar{v}=\mathbf{u}^{T}\mathcal{M}\bar{\mathbf{v}},
\end{equation}
where again $\mathbf{u,v}$ are the components of $u,v$ with respect to the basis $(\alpha_{I},\beta^{J})$, and we use $\mathcal{M}$ as a subscript to distinguish it from the symplectic product. The symmetric matrix $\mathcal{M}$ is called the Hodge matrix and describes the quadratic Hodge norm
\begin{equation}
    \norm{v}_{\mathcal{M}}^2=\langle v,v\rangle_{\mathcal{M}}=\mathbf{v}^{T}\mathcal{M}\bar{\mathbf{v}}.
\end{equation}
From now on let us stick to real forms. Given $u,v\in H^{3}(Y,\mathbb{R})$ and defining $w=*v$, one can combine \eqref{eq:symplectic-product} with \eqref{eq:product-u-v} and write $\mathbf{u}^{T}\mathcal{M}\mathbf{v}=\mathbf{u}^{T}\eta\mathbf{w}$, which implies that $\mathbf{w}=-\eta\mathcal{M}\mathbf{v}$. Representing the Hodge star operator by $\mathcal{M}_{*}$ and using that $Y$ is a Riemannian manifold, it follows that $\mathcal{M}_{*}^2=-\mathbb{I}$ when acting on 3-forms. As a consequence of this, one proves that $\mathcal{M}$ is symplectic 
\begin{equation}
    \mathcal{M}_{*}^2=(-\eta\mathcal{M})^2=\eta\mathcal{M}\eta\mathcal{M}=-\mathbb{I}\hspace{0.8mm}\Longrightarrow \mathcal{M}\eta\mathcal{M}=\eta.
\end{equation}
Let us emphasise at this point that the fact of using the same letter $\mathcal{M}$ for both the Hodge matrix and the ISD matrix in \eqref{eq:third-way-ISD} is not a coincidence, as they are identical. Acting with the Hodge star on both sides of the ISD condition $*G_{3}=iG_{3}$ and picking either the real or the imaginary part of the equation gives $F_{3}-\tau_{1}H_{3}=-\tau_{2}*H_{3}$, which can be written in components as
\begin{equation}
    \mathbf{f}-\tau_{1}\mathbf{h}=\tau_{2}\eta\mathcal{M}\mathbf{h}.
\end{equation}
This is nothing but \eqref{eq:vacua-condition}, where we identify the Hodge matrix with the ISD matrix in \eqref{eq:ISD-matrix}. 

Moving on, any $3$-form $v\in H^{3}(Y,\mathbb{R})$ can be expanded in the aforementioned complex basis as
\begin{equation}
    v=a\Omega+b^{i}D_{i}\Omega+\bar{b}^{i}\overline{D_{i}\Omega}+\bar{a}\bar{\Omega},
\end{equation}
where $a,b^{i}$ are complex coefficients. Using that $*\Omega=-i\Omega$ and $*D_{i}\Omega=i D_{i}\Omega$, the Hodge norm for $v$ reduces to
\begin{equation}
    \norm{v}_{\mathcal{M}}^2=2i\abs{a}^2\langle\Omega,\bar{\Omega}\rangle-2ib^{i}\bar{b}^{j}\langle D_{i}\Omega,\overline{D_{j}\Omega}\rangle,
\end{equation}
which by \eqref{eq:products-forms} takes the form
\begin{equation}
    \norm{v}_{\mathcal{M}}^2=2e^{-K_{\text{cs}}}\left(\abs{a}^2+G_{i\bar{j}}b^{i}\bar{b}^{j}\right).
\end{equation}
The coefficients $a$ and $b^{i}$ are obtained by projecting $v$ onto the $\Omega$ and $D_{i}\Omega$ directions respectively using the symplectic product $\langle\cdot,\cdot\rangle$, so that we can further write
\begin{equation}
    \norm{v}_{\mathcal{M}}^2=2e^{K_{\text{cs}}}\left(\abs{\langle v,\Omega\rangle}^2+G^{i\bar{j}}\langle v,D_{i}\Omega\rangle\overline{\langle v,D_{j}\Omega\rangle}\right).
\end{equation}
By expanding the symplectic products, we read off the Hodge matrix of the quadratic form, which has components
\begin{equation}\label{eq:hodge-matrix-app}
\mathcal{M}_{\alpha\beta}=2e^{K_{\text{cs}}}\text{Re}\left((\eta\mathbf{\Pi})_{\alpha}\overline{(\eta\mathbf{\Pi})}_{\beta}+G^{i\bar{j}}(\eta D_{i}\mathbf{\Pi})_{\alpha}\overline{(\eta D_{j}\mathbf{\Pi})}_{\beta}\right).
\end{equation}
Here the index $\alpha$ runs from $1$ to $2h^{2,1}+2$, having
\begin{equation}
    \norm{v}_{\mathcal{M}}^2=v^{\alpha}\mathcal{M}_{\alpha\beta}v^{\beta},\qquad \mathbf{v}=(v^{\alpha})=\begin{pmatrix}
    v_{1,I}\\
    v_{2}^{I}
\end{pmatrix},
\end{equation}
where the $v$ coefficients are obtained upon expansion in the symplectic basis $v=v_{2}^{I}\alpha_{I}-v_{1,I}\beta^{I}$.

We next aim to derive the two expressions in \eqref{eq:diagonal-components-hodge-matrix} for the diagonal components of the Hodge matrix \eqref{eq:hodge-matrix-app}. Our starting point is the formula \eqref{eq:M-k+1-l+1}
\begin{equation}\label{eq:M-k+1-l+1-app}
    \mathcal{M}_{k+1,l+1}=\frac{2}{V}\left[G^{kl}+\frac{4V_{c}-2C}{V_{c}-2C}x^{k}x^{l}-\frac{6C}{V_{c}-2C}y^{k}y^{l}\right].
\end{equation}
Explicit expressions for $\mathcal{M}_{22}$ and $\mathcal{M}_{33}$ require knowledge of the inverse Kähler components $G^{kl}$. To find these inverse components, we use the Sherman-Morrison formula, which gives the inverse of a matrix with structure $A+\mathbf{u}\mathbf{u}^{T}$, where $A$ is a matrix and $\mathbf{u}$ is a vector. In particular,
\begin{equation}\label{eq:Sherman-Morrison}
    (A+\mathbf{u}\mathbf{u}^{T})^{-1}=A^{-1}-\frac{A^{-1}\mathbf{u}\mathbf{u}^{T}A^{-1}}{1+\mathbf{u}^{T}A^{-1}\mathbf{u}}.
\end{equation}
For our case, we identify
\begin{equation}
    A_{ij}=-\frac{V_{ij}}{4V},\qquad u_{i}=\frac{V_{i}}{2V},
\end{equation}
so that the Kähler metric in \eqref{eq:Kahler-metric-from-kahler-sector} takes the form $G_{ij}=A_{ij}+u_{i}u_{j}$. The inverse of $A$ requires inverting the matrix of volume derivatives $V_{ij}=\partial_{y^{i}}\partial_{y^{j}}V$ which for the geometry with intersection numbers \eqref{eq:kappas_model} reads
\begin{equation}
    V_{ij}=\frac{4}{3}\begin{pmatrix}
        54y^{1}+18y^{2} & 18y^{1}+6y^{2}\\
        18y^{1}+6y^{2} & 6y^{1}\\
    \end{pmatrix},\qquad \text{det}(V_{ij})=-64y^{2}(3y^{1}+y^{2}).
\end{equation}
Applying the formula \eqref{eq:Sherman-Morrison}, we find that the inverse Kähler metric is given by
\begin{equation}
    G^{ij}=-4VV^{ij}-\frac{4VV^{ik}V_{k}V_{l}V^{lj}}{V-V_{p}V^{pq}V_{q}}=-4VV^{ij}+\frac{2Vy^{i}y^{j}}{V_{c}-2C},
\end{equation}
where in the last step we have used the first identity in \eqref{eq:homogeneity-eqs} and $V^{ik}V_{k}=y^{i}/2$, which follows by contracting the second identity in \eqref{eq:homogeneity-eqs} with the inverse matrix components $V^{ij}$. The diagonal components of $G^{ij}$ are
\begin{equation}
    G^{11}=\frac{Vy^{1}}{2y^{2}(3y^{1}+y^{2})}+\frac{2V(y^{1})^2}{V_{c}-2C},\qquad G^{22}=\frac{3V}{2y^{2}}+\frac{2V(y^{2})^2}{V_{c}-2C}.
\end{equation}
Inserting these expressions back in \eqref{eq:M-k+1-l+1-app} and regrouping terms we find the desired Hodge matrix components
\begin{equation}
\begin{aligned}
    \mathcal{M}_{22}=&\frac{4(y^1)^2}{V}+\frac{y^1}{y^2(3y^1+y^2)}+\frac{2}{V}\frac{4V_{c}-2C}{V_{c}-2C}(x^{1})^2,\\\mathcal{M}_{33}=&\frac{4(y^2)^2}{V}+\frac{3}{y^2}+\frac{2}{V}\frac{4V_{c}-2C}{V_{c}-2C}(x^{2})^2.
\end{aligned}
\end{equation}

It remains to prove that the functions $B_{5}$ and $B_{6}$ defined in \eqref{eq:B5-B6-functions} attain their minima when $y^{1}=y^{2}=s$. Since the reasoning behind the proof is similar for both, let us see how this works for $B_{6}$. Recall that
\begin{equation}\label{eq:B6-app}
    B_{6}(y^{1},y^{2})=\frac{3\hspace{0.5mm}\text{max}(y^{1},y^{2})}{\mathcal{M}_{33}},\quad \mathcal{M}_{33}=\frac{4(y^2)^2}{V}+\frac{3}{y^2}+g(V_{c}),\quad g(V_{c})=\frac{2V_{c}-C}{(V_{c}+C)(V_{c}-2C)},
\end{equation}
where we are already fixing $\lvert x^{2}\rvert=\frac{1}{2}$, as explained below equation \eqref{eq:diagonal-components-hodge-matrix} and have introduced the function $g(V_{c})$ for convenience. There are two sectors to be considered. The first one corresponds to $y^{1}\geq y^{2}\geq s$, for which $\text{max}(y^{1},y^{2})=y^{1}$ and so the numerator of $B_{6}$ increases with $y^{1}$. To analyse the denominator, let us keep $y^{2}$ fixed and study its dependence on $y^{1}$. The first term in $\mathcal{M}_{33}$ goes like $1/V$ so is decreasing in $y^{1}$ due to the non-negativity of $\kappa_{ijk}$. The second term is constant in $y^{1}$ and the third one is also decreasing in $y^{1}$, since $g'(V_{c})<0$. Thus, we have an increasing numerator and a decreasing denominator in $y^{1}$, meaning $B_{6}(y^{1},y^{2})\geq B_{6}(y^{2},y^{2})$.

\noindent Along the diagonal $y^{1}=y^{2}=y$, it is not hard to see that $B_{6}$ increases with $y$. By expressing
\begin{equation}
    \frac{1}{B_{6}(y,y)}=\frac{4y}{3(28y^3+C)}+\frac{1}{y^2}+\frac{56y^3-C}{3y(28y^3+C)(28y^3-2C)},
\end{equation}
we see that the RHS decreases with $y$, concluding therefore that $B_{6}(y^{2},y^{2})\geq B_{6}(s,s)$, as desired. Regarding the second sector, one has that $y^{2}\geq y^{1}\geq s$. Then, $\text{max}(y^{1},y^{2})=y^{2}$ and
\begin{equation}\label{eq:inverse-B6-app}
    \frac{1}{B_{6}(y^{1},y^{2})}=\frac{4y^{2}}{3V}+\frac{1}{(y^2)^2}+\frac{g(V_{c})}{3y^{2}}.
\end{equation}
As we did before, holding $y^{2}$ fixed while varying $y^{1}$, we see that the RHS is decreasing in $y^{1}$. This implies that $B_{6}(y^{1},y^{2})\geq B_{6}(s,y^{2})$. Lastly, we simply need to check that when varying $y^{2}$ while fixing $y^{1}=s$, the RHS in \eqref{eq:inverse-B6-app} decreases as $y^{2}$ increases. Clearly, the second term is decreasing in $y^{2}$. The same happens for the third term since both $g(V_{c})$ and $1/y^{2}$ are positive and decreasing. To understand the behaviour of the first term, we look at its derivative with respect to $y^{2}$. That is,
\begin{equation}
    \frac{\partial}{\partial y^{2}}\left(\frac{y^{2}}{V(s,y^{2})}\right)=\frac{12s^{3}-4s(y^{2})^2+C}{V^{2}},
\end{equation}
where $V(s,y^{2})=12s^{3}+12s^{2}y^{2}+4s(y^{2})^2+C$. In this case, at $y^{2}=s$, the derivative is positive, and as $y^{2}$ increases the term $-4s(y^{2})^2$ will make the numerator smaller, to the point that it becomes negative for $y^{2}>(3s^2+\frac{C}{4s})^{1/2}$. However, despite the fact that the derivative is positive for $y^{2}\in[s,(3s^2+\frac{C}{4s})^{1/2}]$, one easily checks that along this region the first term in the RHS of \eqref{eq:inverse-B6-app} grows much more slowly than the second term shrinks. The rate at which the first term increases is bounded by
\begin{equation}
    \frac{\partial}{\partial y^{2}}\left(\frac{4y^{2}}{3V(s,y^{2})}\right)=\frac{4}{3}\frac{12s^{3}-4s(y^{2})^2+C}{V^{2}}<\frac{4}{3}\frac{12s^{3}+C}{(28s^{3}+C)^2}<\frac{4}{3(28s^{3}+C)}<\frac{1}{21s^{3}}\simeq\frac{0.05}{s^{3}},
\end{equation}
whereas for the second term we have that
\begin{equation}
    \frac{\partial}{\partial y^{2}}\left(\frac{1}{(y^{2})^2}\right)=-\frac{2}{(y^{2})^3}<-\frac{2}{\left(3s^2+\frac{C}{4s}\right)^{3/2}}<-\frac{2}{8.94s^{3}}\simeq-\frac{0.22}{s^{3}}.
\end{equation}
In the last inequality we considered that $s>1$. Hence, $B_{6}(s,y^{2})$ is an increasing function in $y^{2}$, having $B_{6}(s,y^{2})\geq B_{6}(s,s)$. Combining both sectors proves $B_{6}(y^{1},y^{2})\geq B_{6}(s,s)$ for all $y^{1},y^{2}\geq s$. 

To conclude this appendix, another useful formula, used to derive the bounds \eqref{eq:hodge-bound-H3} and \eqref{eq:hodge-bound-F3} is the Cauchy-Schwarz inequality with underlying Hodge product. Given two forms $u,v\in H^{3}(Y,\mathbb{R})$, this inequality tells us that
\begin{equation}\label{eq:cauchy-schwartz-bound}
    \langle u,v\rangle_{\mathcal{M}}^2\leq \langle u,u\rangle_{\mathcal{M}}\langle v,v\rangle_{\mathcal{M}}.
\end{equation}
Now, taking $\mathbf{u}=\mathcal{M}^{-1}\mathbf{e}_{\alpha}$, the product \eqref{eq:product-u-v} reduces to
\begin{equation}
    \langle u,v\rangle_{\mathcal{M}}=(\mathcal{M}^{-1}\mathbf{e}_{\alpha})^{T}\mathcal{M}\mathbf{v}=\mathbf{e}_{\alpha}^{T}\mathbf{v}=v^{\alpha},
\end{equation}
while the self-product of $u$ picks up components of the inverse Hodge matrix as
\begin{equation}
    \langle u,u\rangle_{\mathcal{M}}=(\mathcal{M}^{-1}\mathbf{e}_{\alpha})^{T}\mathcal{M}(\mathcal{M}^{-1}\mathbf{e}_{\alpha})=\mathcal{M}^{-1}_{\alpha\alpha}=\mathcal{M}^{\alpha\alpha}.
\end{equation}
The bound in \eqref{eq:cauchy-schwartz-bound} now translates into
\begin{equation}\label{eq:cauchy-schwartz-bound-2}
    \norm{v}_{\mathcal{M}}^2\geq \frac{(v^{\alpha})^2}{\mathcal{M}^{\alpha\alpha}},
\end{equation}
where of course no summation over $\alpha$ is taken.

\bibliographystyle{utphys}
\bibliography{biblio.bib}

\end{document}